\documentclass[lettersize,journal]{IEEEtran}
\usepackage{amsmath,amsfonts}
\usepackage{array}
\usepackage[caption=false,font=normalsize,labelfont=sf,textfont=sf]{subfig}

\usepackage{textcomp}
\usepackage{stfloats}
\usepackage{url}
\usepackage{verbatim}
\usepackage{graphicx}
\usepackage{cite}
\usepackage{physics}
\usepackage{amssymb}
\usepackage{orcidlink}
\usepackage{booktabs, multirow}
\usepackage[table,svgnames]{xcolor}
\usepackage{hyperref}
\allowdisplaybreaks
\definecolor{panblue}{RGB}{20, 66, 129}
\usepackage{amsthm}
\newtheoremstyle{custom}
  {1.5ex}{1ex}{\itshape}{}{\bfseries}{}{0pt}%
  {\thmname{#1}\ \thmnumber{#2}\thmnote{ (#3)}: }
\theoremstyle{custom}
\newtheorem{theorem}{Theorem}
\newtheorem{definition}{Definition}
\newtheorem{property}{Property}
\newcommand{\out}[1]{\ensuremath{\ket{#1}\!\!\bra{#1}}}
\newcommand{\xx}[1]{\texttt{#1}}                                                   
\newcommand{\strat}{\ensuremath{\boldsymbol{\lambda}}}                             
\newcommand{\outa}{\ensuremath{\mathbf{a}}}                                        
\newcommand{\stratrest}[1]{\ensuremath{\boldsymbol{\lambda}}^{(\text{rest},#1)}}   
\newcommand{\cond}{\ensuremath{\,|\,}}                                             
\newcommand{\eqalign}[1]{\begin{equation}\begin{aligned} #1 \end{aligned}\end{equation}}

\usepackage{tcolorbox}
\tcbuselibrary{skins,breakable}
\usetikzlibrary{shadings,shadows}

    {\endtcolorbox}

\newenvironment{sumblocknt}[1]{%
    \tcolorbox[beamer, 
    noparskip,breakable,
    colback=white!99!MidnightBlue,
    colframe=black!10!panblue,%
    colbacktitle=panblue,
    coltitle=white,
    colbacklower=white,%
    boxrule=1pt,
    enhanced jigsaw,
    no shadow,
    toptitle=0pt,
    bottomtitle=2pt,
    scale=1
    ]}%
    {\endtcolorbox}

\begin{document}

\title{Linear Program Witness for Network Nonlocality in Arbitrary Networks \thanks{A preliminary version of this work appeared in the Proceedings of QuNet'25 presenting a general methodology \cite{hayes2025linear}.
}}

\author{Salome Hayes-Shuptar \orcidlink{0009-0001-9588-5440}, Daniel Bhatti \orcidlink{0000-0002-8762-3053}, Ana Belen Sainz \orcidlink{0000-0003-3123-8436}, David Elkouss \orcidlink{0000-0003-2023-2768}
\thanks{Salome Hayes-Shuptar is with the Networked Quantum Devices Unit, Okinawa Institute of Science and Technology, Okinawa, Japan (e-mail: s.hayes@oist.jp).}
\thanks{Daniel Bhatti is with the Networked Quantum Devices Unit, Okinawa Institute of Science and Technology, Okinawa, Japan (e-mail: 
daniel.bhatti@fau.de).}
\thanks{Ana Belen Sainz is with the International Centre for Theory of Quantum Technologies, University of  Gda{\'n}sk, 80-309 Gda{\'n}sk, Poland (email: ana.sainz@ug.edu.pl). She is also with the Theoretical Sciences Visiting Program, Okinawa Institute of Science and Technology, Okinawa, Japan.}
\thanks{David Elkouss is with the Networked Quantum Devices Unit, Okinawa Institute of Science and Technology, Okinawa, Japan (e-mail: 
david.elkouss@oist.jp).}
}

\maketitle

\begin{abstract}
Network nonlocality extends Bell nonlocality to settings with multiple independent sources and parties. Certifying it in quantum information processing tasks requires suitable witnesses. However, in contrast to local correlations, the set of network-local correlations is non-convex. This non-convexity makes certifying network nonlocality a highly non-trivial task. Existing approaches involve leveraging network-specific properties, or inflation-based methods whose constraints grow combinatorially in the number of local variables. In this work, we introduce a linear programming witness for network nonlocality built from five classes of linear constraints. These classes are network-agnostic, although the explicit forms of the constraints must be tailored to a specific network's structure. We use the procedure to construct network nonlocality witnesses for a family of ring networks and certify network nonlocality for a concrete example, relying only on observed probabilities and a tunable experimental parameter. Our work advances the search for efficient witnesses to certify network nonlocality across diverse quantum network architectures.
\end{abstract}

\section{Introduction}
In classical theory, we assume local realism and measurement independence \cite{brunner2014bell}. Local realism means measurement outcomes are fixed by properties in their past light-cones, while measurement independence means measurement settings are chosen freely and uncorrelated from those properties. These assumptions form the basis of local models, in which all correlations arise from local ``hidden" variables. In 1964, Bell derived an inequality that correlations explained by local variables (LV) must satisfy, and showed that quantum correlations may violate it\cite{bell1964einstein}. If the statistical data (e.g., correlations) violates a Bell inequality, meaning at least one of the classical assumptions failed, we say it is \textit{nonlocal}. Notably, correlations arising from quantum systems can violate Bell inequalities, while correlations from classical systems always satisfy them. This fact has been used to develop many practical applications for nonlocality in quantum information processing tasks, including certifying entanglement and randomness, and establishing secure cryptographic keys \cite{barrett2005no, pironio2010random}. Nonlocality is a particularly useful tool as it only relies on the measurement statistics of a system, without needing to trust the devices themselves. This property allows nonlocality to serve as a resource for device-independent (DI) protocols \cite{acin2007device}.

Quantum networks are systems of multiple independent sources distributing information to spatially separated parties \cite{kimble2008quantum,wehner2018quantum}. Their additional structure enables phenomena beyond single-source setups, such as entanglement swapping for linking previously unentangled nodes \cite{zukowski1993event}. The study of quantum networks has also motivated a new type of nonlocality, called \textit{network-nonlocality}, which supplements the usual Bell assumptions with conditions stemming from the specific structure of the sources in the network: in a network-local model, one may no longer have a single source connecting all the parties, but rather a collection of sources (whose local variables are independent of the other sources) each connected to specific parties. When the correlations of a system cannot be reproduced by a model that satisfies local realism, measurement independence, and extra conditions which we refer to as \textit{source independence}, we say the system exhibits network-nonlocality \cite{tavakoli2022bell}.

Beyond its significance in quantum foundations, there is also the question of whether network nonlocality can be used as a resource for tasks in quantum protocols, similar to how Bell nonlocality is used for DI certification. This requires furthering our understanding of its prevalence in practical networks, its robustness to experimental constraints, and how we can operationally quantify it \cite{chen2021device}. Investigating it is a difficult task due to the fact that the set of network-local correlations is non-convex, whereas standard local correlations form a convex set amenable to (more) efficient convex optimization tools \cite{brito2018quantifying}\footnote{We note convex optimization for standard Bell scenarios is inefficient and only tractable for a modest number of inputs\cite{avis2005two}.}. Thus, the first step is to establish more systematic techniques for witnessing network nonlocality. Current methods involve nonlinear inequalities derived on a network-specific basis, such as chain and ring networks with bipartite sources, or inflation-based methods whose constraints grow combinatorially in the number of local variables \cite{wolfe2021quantum, pozas2019bounding}.

While numerical witnesses for arbitrary network topologies have yet to be developed, Ref.~\cite{abiuso2022single} demonstrated a linear program (LP) witness for network nonlocality in an $N$-party ring network with $N$ independent single-photon sources, tunable beamsplitters, and click/no-click detectors. Building on their work, we present a general linear programming approach for witnessing network nonlocality by outlining five sufficient classes of linear constraints on an auxiliary distribution related to the measurement statistics. The constraint classes themselves are network-agnostic, however the form of some constraint classes will depend on the specific network structure. In particular, classes 1 and 2 are generic for any network, while classes 3-5 exploit network-specific structure. Nevertheless, an LP witness for network nonlocality is generally preferable to semi-definite relaxations, which are sometimes intractable~\cite{vandenberghe1996semidefinite}.

In this work, we demonstrate our method with a concrete example for a family of ring networks with arbitrary source-to-party configurations. As an example, we use our linear program to certify the network-nonlocality of correlations arising from a similar photonic setup to that in Ref.~\cite{abiuso2022single}, and obtain results in terms of the transmissivity of the beamsplitters at each party in the ring network. We further discuss how to derive the network-specific constraints leveraging the physical properties of such a photonic system.

We organize the remainder of this paper as follows. Sec.~\ref{sec:background} reviews definitions and Sec.~\ref{sec:general_lp_procedure} outlines the general LP procedure. Sec.~\ref{sec:ring_network_construction} introduces the networks considered in this work and Sec.~\ref{sec:lp_procedure_ring_network} derives the corresponding analytical form of the LP constraints. Sec.~\ref{sec:6p4s_network} illustrates our method for a concrete network scenario and we conclude with a discussion of the results in Sec.~\ref{sec:conclusion}.

\section{Background}\label{sec:background}
\subsection{Standard Bell Nonlocality}\label{subsec:standard_bell_nonlocality}
\begin{figure}
    \centering
    \includegraphics[width=0.95\linewidth]{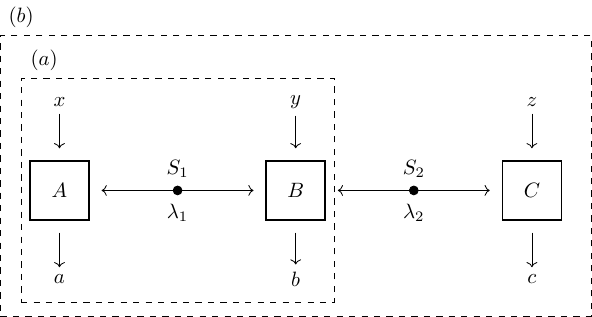}
    \caption{(a) Bipartite Scenario with one source $S_1$ characterized by one LV $\lambda_1$, distributing states to two parties, $A$ and $B$. (b) Bilocal scenario with two sources $S_1$ and $S_2$, characterized by two LVs $\lambda_1$ and $\lambda_2$, distributing states to three parties $A$, $B$, and $C$. The parties receive inputs $(x,y,z)$, determining their measurement settings and return measurement outcomes $(a,b,c)$.}
    \label{fig:bilocal_network}
\end{figure}

Consider the scenario in Fig.~\ref{fig:bilocal_network}(a). A source $S_1$ prepares a system in a state and distributes it to two spatially separated parties, $A$ and $B$. Each party chooses from a finite set of measurement settings, chosen randomly by inputs $x$ and $y$, performs their measurement on their share of the system, and returns outputs $a$ and $b$. The object of study is the correlations $p(a,b \cond x,y)$.

Bell locality is defined for correlations which can be characterized by one global LV \cite{clauser1969proposed}. Under the assumption of a classical model, all correlations between $A$ and $B$ arise due to an LV $\lambda_1$, at the source. We will use `local' as a shorthand for `admits a classical model'. We say the system is local if the joint distribution of the measurement outcomes can be decomposed as
\begin{align}\label{eqn:single_system_locality}
    p(a,b\cond x,y)=\int d\lambda_1\: p_1(\lambda_1)\:p_a(a\cond x,\lambda_1)\: p_b(b\cond y,\lambda_1),
\end{align}
where $p_1(\lambda_1)$ is the distribution of the LV, and $p_a(a\cond x,\lambda_1)$ and $p_b(b\cond y,\lambda_1)$ are the response functions of $A$ and $B$, respectively. We call $p(a,b\cond x,y)$ the \textit{behavior} of the system. If the behavior does not factorize as in Eqn.~\eqref{eqn:single_system_locality}, we say the system is nonlocal. These definitions can be naturally generalized to larger single-source systems.

The set of all local behaviors $\mathcal{L}$, is a convex polytope~\cite{goh2018geometry}. Its vertices correspond to behaviors with deterministic response functions, meaning $\lambda_1$ maps inputs to a fixed output. A specific choice of mapping between the inputs and outputs is called a \textit{strategy}. Thus each vertex represents a behavior with a deterministic strategy. The facets of $\mathcal{L}$ are hyperplanes corresponding to linear Bell inequalities, which any local behavior must satisfy. These inequalities form experimentally friendly witnesses for standard Bell nonlocality, as they only involve linear combinations of measurement correlators. However, we can also witness nonlocality by constructing a membership test for the observed behavior $p(a,b\cond x,y)$ in $\mathcal{L}$ \cite{kaszlikowski2000violations}. We cast this as a feasibility test. Since $\mathcal{L}$ is convex, this test is a special case of convex optimization which checks for the existence of at least one point inside the polyhedron defined by a set of linear constraints.

We illustrate this for the scenario in Fig.~\ref{fig:bilocal_network}(a), although it can be generalized for other single-source systems. First, we enumerate all deterministic strategies and label each strategy by the pair of bit-strings $\bar{\alpha}\bar{\beta}$, where
\begin{align}\label{eqn:strategy_index_bitstrings}
    \bar{\alpha}=\alpha_0\alpha_1, && \bar{\beta}=\beta_0\beta_1,
\end{align}
encode the outputs at $A$ and $B$ for inputs $x,y\in\{0,1\}$ (e.g. $\alpha_0$ indicates the output of $A$ when it receives $x=0$ and $\beta_0$ indicates the output of $B$ when it receives $y=0$). For outputs $a,b\in\{0,1\}$, the string $\bar{\alpha}\bar{\beta}=0110$ means $\lambda_1$ assigns the input-output maps as $0\mapsto 0$ and $1\mapsto 1$ at $A$, and $0\mapsto 1$ and $1\mapsto 0$ at $B$. For binary inputs and outputs there are $2^2\times 2^2=16$ such strategies and thus the local set has 16 vertices. Any local behavior of the system can be written as a convex mixture over the deterministic strategies~\cite{brunner2014bell},
\begin{align}
    p(a,b\cond x,y) &=\sum_{\bar{\alpha}\bar{\beta}} w_{\bar{\alpha}\bar{\beta}} \:\delta_{a,\alpha_x}\:\delta_{b,\beta_y},\label{eqn:standard_lp_constraint1}\\
    &w_{\bar{\alpha}\bar{\beta}}=\int\displaylimits_{\Lambda_1^{\bar{\alpha}\bar{\beta}}}d\lambda_1\: p_1(\lambda_1)\geq 0,\label{eqn:standard_lp_constraint2}\\
    &\sum_{\bar{\alpha}\bar{\beta}}w_{\bar{\alpha}\bar{\beta}}=1,\label{eqn:standard_lp_constraint3}
\end{align}
where $\Lambda_1^{\bar{\alpha}\bar{\beta}}$ indicates the domain of $\lambda_1$ that produces strategy $\bar{\alpha}\bar{\beta}$ (i.e. $\Lambda_1^{\bar{\alpha}\bar{\beta}}=\{\lambda_1\mid \lambda_1\rightarrow \bar{\alpha}\bar{\beta}\}$). The weights $w_{\bar{\alpha}\bar{\beta}}$ indicate the probability of strategy $\bar{\alpha}\bar{\beta}$ occurring. The subscripts $\alpha_x$ and $\beta_y$ in Eqn.~\eqref{eqn:standard_lp_constraint1} indicate the strategy's output for a given input (e.g. $\alpha_x\in\{\alpha_0,\alpha_1\}$ and $\beta_y\in\{\beta_0,\beta_1\}$). Eqns.~\eqref{eqn:standard_lp_constraint1}-\eqref{eqn:standard_lp_constraint3}, form a feasibility test for a set of weights $w_{\bar{\alpha}\bar{\beta}}$. Since all three are linear, the test is an LP with no objective function.

If weights satisfying the constraints exist, we say the LP has certified a local model explanation of the system. If the constraints cannot be satisfied, the LP is infeasible and certifies nonlocality.

\subsection{Network Nonlocality}\label{subsec:network_nonlocality}

Consider the network in Fig.~\ref{fig:bilocal_network}(b), also known as the bilocal scenario \cite{branciard2012bilocal}. Two independent sources each ($S_1$ and $S_2$) prepare a system in a state and distribute it to three spatially separated parties $A$, $B$, and $C$. The parties have inputs $x$, $y$, and $z$ which determine their measurement settings, and return outputs $a$, $b$, and $c$ respectively. The object of study is the correlations $p(a,b,c\cond x,y,z)$. 

Network-locality is defined for correlations which are characterized by multiple independent LVs. Under the assumption of a classical model, all correlations here arise from two LVs, $\lambda_1$ at source $S_1$ and $\lambda_2$ at source $S_2$. We say the system is network-local if its behavior admits the following decomposition:
\eqalign{
    p(a,b,c\cond &x,y,z) = \int d\lambda_1\:p_1(\lambda_1)\!\int d\lambda_2\:p_2(\lambda_2)\\
    &\times p_a(a\cond x,\lambda_1)\:p_b(b\cond y,\lambda_1,\lambda_2)\:p_c(c\cond z,\lambda_2)\,,
    \label{eqn:bilocal_locality}
}
where $p_m(\lambda_m)$ are normalized distributions. Unlike the standard locality condition in Eqn.~\eqref{eqn:single_system_locality}, the network-local condition in Eqn.~\eqref{eqn:bilocal_locality} requires the distributions of the independent LVs also factorize. If the joint statistics of a network cannot be written as in Eqn.~\eqref{eqn:bilocal_locality}, we say the system exhibits network nonlocality. These definitions can be naturally generalized to larger networks.

The set of all network-local behaviors $\mathcal{NL}$ and that of local ones $\mathcal{L}$ share the same extremal points and obey $\mathcal{NL}\subset \mathcal{L}$, so that $\mathcal{L}$ is the convex hull of $\mathcal{NL}$ \cite{fritz2012beyond}. However, a consequence of imposing the extra assumption of source independence is that $\mathcal{NL}$ is non-convex and its facet inequalities are nonlinear. The membership test for the behavior $p(a,b,c\cond x,y,z)$ in $\mathcal{NL}$ is no longer a convex feasibility problem, but a non-convex polynomial optimization problem \cite{fritz2012beyond}. Nevertheless, it is instructive to outline how to construct this test as a comparison for our later linearized approach.

We consider the scenario in Fig.~\ref{fig:bilocal_network}(b), but this construction can be generalized for arbitrary networks. Since $\mathcal{NL}\subset \mathcal{L}$, we decompose the behavior as if it were in $\mathcal{L}$, then impose source independence to correctly restrict to $\mathcal{NL}$. Again, we enumerate all deterministic strategies of input to output mapping for each party. Using the same convention as in Eqn.~\eqref{eqn:strategy_index_bitstrings}, we define $\bar{\alpha}$, $\bar{\beta}$, and $\bar{\gamma}$ as the indices of the strategies for each party, $A$, $B$, and $C$, respectively. Then we write the behavior as a weighted sum over deterministic response functions,
\begin{align}
    p(a,b,c\cond x,y,z)= \sum_{\bar{\alpha}\bar{\beta}\bar{\gamma}} w_{\bar{\alpha}\bar{\beta}\bar{\gamma}} \:\delta_{a,\alpha_x}\:\delta_{b,\beta_y}\:\delta_{c,\gamma_z},\\
    w_{\bar{\alpha}\bar{\beta}\bar{\gamma}}=\int\displaylimits_{\Lambda_{1,2}^{\bar{\alpha}\bar{\beta}\bar{\gamma}}}d\lambda_1 d\lambda_2\: p_1(\lambda_1)\: p_2(\lambda_2)\geq 0,\label{eqn:network_lp_constraint2}\\
    \sum_{\bar{\alpha}\bar{\beta}\bar{\gamma}}w_{\bar{\alpha}\bar{\beta}\bar{\gamma}}=1,
\end{align}
where $\Lambda_{1,2}^{\bar{\alpha}\bar{\beta}\bar{\gamma}}$ indicates the joint domain of the LV pair $(\lambda_1,\lambda_2)$ which produce strategy $\bar{\alpha}\bar{\beta}\bar{\gamma}$ (i.e. $\Lambda_{1,2}^{\bar{\alpha}\bar{\beta}\bar{\gamma}}=\{(\lambda_1,\lambda_2)\mid \lambda_1\rightarrow \bar{\alpha}, \lambda_2\rightarrow \bar{\gamma}, (\lambda_1,\lambda_2)\rightarrow \bar{\beta}\}$), and $w_{\bar{\alpha}\bar{\beta}\bar{\gamma}}$ indicates the probability of strategy $\bar{\alpha}\bar{\beta}\bar{\gamma}$ to occur. Note that Eqn.~\eqref{eqn:network_lp_constraint2} is nonlinear, meaning the three equations form a non-convex feasibility test for a set of weights $w_{\bar{\alpha}\bar{\beta}\bar{\gamma}}$. Infeasibility certifies network nonlocality. However, this presents a program which is computationally expensive at best, and intractable at worst. Indeed, global non-convex polynomial optimization is NP-hard~\cite{murty1985some}.

\section{General Linear Program Procedure}\label{sec:general_lp_procedure}
We now discuss the general procedure for constructing an LP witness for network nonlocality, which we introduced in the previous work \cite{hayes2025linear}. It follows three main steps:
\begin{enumerate}
    \item Enumerate deterministic LV strategies.
    \item Isolate an amenable outcome subset.
    \item Construct the constraints for the LP.
\end{enumerate}
For \textit{Step 3}, we separate the constraints into five classes. Infeasibility in the resulting LP is a sufficient test to rule out a network-local model. In principle, this method only assumes the existence of measurement statistics, meaning it is applicable for arbitrary networks. We note for network scenarios, this corresponds to the memoryless regime \cite{Weilenmann2025}. In practice, finding classes 3-5 for the LP requires knowledge of a given network's properties.

For now, consider an arbitrary network with $N$ parties and $M$ independent sources. Let
\begin{align}
    \mathcal{O}=\{\outa\}, && \text{with }\outa=(a_1,\dots,a_n,\dots,a_N),
\end{align}
where $a_n$ indicates the output of party $A_n$, be the set of all outcomes in the network that could potentially be observed in the experiment. For example, a photonic setup with photon number conservation would require the number of detector clicks for each outcome string $(a_1,\dots,a_n,\dots,a_N)$ be compatible\footnote{If the detectors are not photon-number-resolving, then the number of clicks can be smaller than the number of photons. } with the number of photons in the network. 

\begin{figure}
    \centering
    \includegraphics[width=0.95\linewidth]{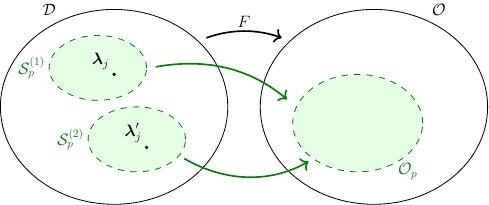}
    \caption{$\mathcal{D}$ denotes the LV domain and $\mathcal{O}$ denotes the output space, which are related via the map $F(\mathcal{D})\mapsto \mathcal{O}$. $\mathcal{S}_p^{(1)}$ and $\mathcal{S}_p^{(2)}$ denote subregions in the LV space, $\mathcal{O}_p=F(\mathcal{S}_p^{(1)}\sqcup \mathcal{S}_p^{(2)})$ denotes a subregion in output space, and $\strat_j$ and $\strat_j'$ denote specific strategies in $\mathcal{S}_p^{(1)}$ and $\mathcal{S}_p^{(2)}$ respectively.}
    \label{fig:domain_mappings}
\end{figure}
We now discuss the construction of a network-local model for the scenario. In such a model, each source $S_m$ carries its own independent LV $\lambda_m$, drawn from a domain $\Lambda_m$. The joint LV domain is then the Cartesian product
\begin{align}
    \mathcal{D}&= \prod_{m=1}^M \Lambda_m = \{\strat=(\lambda_1,\dots,\lambda_M)\cond \lambda_m\in\Lambda_m\},
\end{align}
where the vector $\strat$ represents a deterministic strategy in LV space. Let $F$ be the map $\mathcal{D}\mapsto \mathcal{O}$, which maps every deterministic strategy $\strat$ to its assigned global outcome $\outa$ in $\mathcal{O}$, as shown in Fig.~\ref{fig:domain_mappings}.

For \textit{Step 2}, consider a subset $\mathcal{O}_S\subset \mathcal{O}$, and define $\mathcal{S}=F^{-1}(\mathcal{O}_S)$. At this moment, we leave $\mathcal{O}_S$ to be a fixed-but-arbitrary set, and later we will see how choosing it in a clever way allows one to define meaningful constraints for the LP we are constructing.

Let $p(\outa)$ denote the joint distribution over $\mathcal{O}$ observed in the experiment, where we omit input dependence\footnote{Any system, single or multi-source, can be transformed to one with fixed measurement settings by mapping the inputs of the parties to outputs of new parties \cite{boreiriBell2025}.}. If $\{p(\outa)\}_{\outa \in \mathcal{O}}$ admits a network-local model, it can be written as 
\begin{align}\label{eqn:mu_def}
    p(\outa) = \sum_{\strat \in \mathcal{D}
    } \mu(\outa,\strat)\,,
\end{align}
where the probability distribution $\{\mu(\outa,\strat)\}_{\outa \in \mathcal{O},\strat \in \mathcal{D}}$ satisfies extra constraints due to network-local requirements (such as source independence), which will play an important role below. 

We now define an auxiliary probability distribution over the restricted subsets $\{q(\outa,\strat)\}_{\outa \in \mathcal{O}_S,\strat \in \mathcal{S}}$ which we relate to $\mu$ via
\begin{align} \label{eq:qandmu}
    q(\outa,\strat) \equiv \frac{\mu(\outa,\strat)}{\sum_{\outa' \in \mathcal{O}_S,\strat' \in \mathcal{S}} \mu(\outa',\strat')} \,.
\end{align}
We would like to write constraints on $q(\outa,\strat)$ which hold whenever $p(\outa)$ is network-local. However, $\mu(\outa,\strat)$ is not accessible from experimental data, and thus such constraints on $q(\outa,\strat)$ must be specified in terms of $p(\outa)$ alone. In what follows, we will list linear constraints on $q(\outa,\strat)$ that define the LP over $q(\outa,\strat)$ (our main result): if the LP cannot find a $q(\outa, \strat)$ satisfying the constraints and is compatible with the observed statistical data (i.e. $p(\outa)$), then we conclude $p(\outa)$ is network nonlocal. We highlight infeasibility in the LP we define is a \textit{sufficient} condition  for network nonlocality, but not a \textit{necessary} one. That is, a solution $q(\outa, \strat)$ for the LP does not imply $p(\outa)$ is network-local. The decision variables in the LP are the joint probabilities,
\begin{align}
    q(\outa,\strat), && \forall\outa\in \mathcal{O}_S,\: \strat\in\mathcal{S}.
\end{align}
We now introduce the five classes of linear constraints forming the basis for the LP, which is a feasibility test for $q(\outa,\strat)$. We will comment on how our LP compares to other numerical tools in Sec.~\ref{sec:conclusion}. 

\subsection*{\textbf{Constraint Class 1: Distribution Validity}}
These constraints are generic for any network and ensure that $q(\outa,\strat)$ is a valid probability distribution over the variables $\{\outa,\strat\}$, and correspond to non-negativity and normalization:
\eqalign{
   q(\outa, \strat)&\geq 0, \qquad \forall \outa \in\mathcal{O}_S, \forall \strat\in\mathcal{S},\\
   \sum_{\outa\in\mathcal{O}_S}\sum_{\strat\in\mathcal{S}}q(\outa,\strat)&=1.
   \label{eqn:constraints1}
}

\subsection*{\textbf{Constraint Class 2: Marginal Agreement}}
These constraints relate  $q(\outa,\strat)$ to the observed statistical data $p(\outa)$. For this, let us start from Eqn.~\eqref{eq:qandmu} and take the marginal of $q$ over $\strat$:
\begin{align}
    q(\outa)&=\sum_{\strat\in\mathcal{S}} q(\outa,\strat) \\
    &= \frac{1}{\sum_{\outa' \in \mathcal{O}_S,\strat' \in \mathcal{S}} \mu(\outa',\strat')} \sum_{\strat\in\mathcal{S}} \mu(\outa,\strat) \\
    &= \frac{p(\outa)}{\sum_{\outa'\in\mathcal{O}_S}p(\outa')}\,,
\end{align}
since $\mu(\outa,\strat) = 0$ whenever $\outa \in \mathcal{O}\setminus \mathcal{O}_S$ and $\strat \in \mathcal{S}$, and also $\mu(\outa,\strat) = 0$ whenever $\outa \in \mathcal{O}_S$ and $\strat \in \mathcal{D} \setminus \mathcal{S}$\,. This imposes the following constraints, which are generic for any network:
\begin{align}
    q(\outa)&=\sum_{\strat\in\mathcal{S}} q(\outa,\strat), \label{eqn:qa_marginal} \\
    q(\outa)&=\frac{p(\outa)}{\sum_{\outa'\in\mathcal{O}_S}p(\outa')}, && \forall \outa\in\mathcal{O}_S.\label{eqn:constraints2}
\end{align}

\subsection*{\textbf{Constraint Class 3: Strategy Distribution}}
These constraints pertain to the marginal distribution $q(\strat)$, which when $\{p(\outa)\}_{\outa \in \mathcal{O}}$ is network-local can be written as 
\begin{align}\label{eqn:q_lambda}
    q(\strat) = \sum_{\outa \in \mathcal{O}_S} q(\outa, \strat) =  \frac{\mu(\strat)}{\sum_{\strat' \in \mathcal{S}} \mu(\strat') } \,,
\end{align}
following Eqn.~\eqref{eq:qandmu}. 
The explicit form of $\{\mu(\strat)\}_{\strat \in \mathcal{D}}$ for a hypothetical network-local model is of course unknown, so the trick here is to relate $\{\mu(\strat)\}_{\strat \in \mathcal{D}}$ to the observed statistics $\{p(\outa)\}_{\outa \in \mathcal{O}}$ whenever possible.

\subsection*{\textbf{Constraint Class 4: Conditional Independence}}
These constraints translate locality properties of the $\mu(\outa,\strat)$ pertaining to a hypothetical network-local $p(\outa)$ into constraints for $q(\outa,\strat)$. More specifically, we leverage conditional independence to write the single-party marginal
\begin{align}\label{eqn:spm}
    p_n(a) := \sum_{\outa \in \mathcal{O} \,:\, a_n = a } p(\outa), 
\end{align}
such that it can only depend on the value of the hidden variables from the source party $A_n$ is connected to. How to translate this into constraints for $q(\outa,\strat)$ depends on the network.

\subsection*{\textbf{Constraint Class 5: Domain Asymmetry}}
These constraints aim to capture the `bias' in the sampling of local variables needed to reproduce the observed data, as we explain below.  

We define the constraints in this class leveraging the fact that the same outcome could arise from disjoint parts of the LV domain. To motivate the definition, consider Fig.~\ref{fig:domain_mappings} again. Let $\mathcal{O}_p=\{\outa_i\}\subseteq \mathcal{O}_S$ be a subset of outcomes with pre-image $\mathcal{S}_p=\mathcal{S}_p^{(1)}\sqcup \mathcal{S}_p^{(2)}$. If the statistical data admits a network-local model, every time the system outputs $\outa_i\in\mathcal{O}_p$, it must ``sample" its strategy from the two possible regions $\mathcal{S}_p^{(1)}$ or $\mathcal{S}_p^{(2)}$. We can compute the asymmetry, or bias, in the regions as
\begin{align} \label{eq:16}
    \Delta_{\mathcal{O}_p}'&\!\!=q\big(\strat\in\mathcal{S}_p^{(1)}\mid \outa\in\mathcal{O}_p\big)-q\big(\strat\in\mathcal{S}_p^{(2)}\mid \outa\in\mathcal{O}_p\big),\\
    \Delta_{\mathcal{O}_p}&\!\!=\Delta_{\mathcal{O}_p}'q\big(\outa\in\mathcal{O}_p\big). \label{eqn:general_domain_constraint}
\end{align}

Then to specify these constraints, one needs to compute the analytical forms of Eqns.~\eqref{eq:16} and \eqref{eqn:general_domain_constraint} from our observed statistics $\{p(\outa)\}_{\outa \in \mathcal{O}}$. This depends further on the specific form of the network.

Intuitively, the domain asymmetry constraints aim to capture the idea that some network-nonlocal correlations will require more bias in the strategy weights than allowed by Eqn.~(\ref{eqn:general_domain_constraint}). We can understand this as the parties needing extra information or ``coordination" not permitted by independent LVs, in order to agree with the observed statistics.

\section{Ring Network Construction}\label{sec:ring_network_construction}
\begin{figure}
    \centering
    \includegraphics[width=0.75\linewidth]{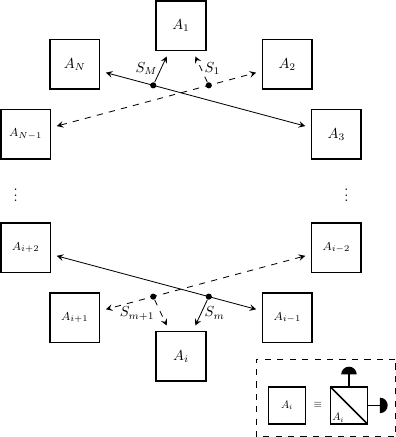}
    \caption{Ring network with $M$ sources, $S_1,\dots,S_M$, distributing tripartite single-photon W states to $N$ parties, $A_1,\dots,A_N$. Each party is equipped with a tunable beamsplitter and a non-photon-number-resolving detector in each optical mode. Each source $S_m$ has an associated LV $\lambda_m$. For this specific source-to-party configuration, odd-numbered sources signal to the next-nearest-neighbor to the left of their middle party and the nearest-neighbor to the right of their middle party, and vice versa for even-numbered sources.}
    \label{fig:ring_network_wstates}
\end{figure}

We now introduce the family of ring networks through which we will demonstrate how to construct and apply our LP. Fig.~\ref{fig:ring_network_wstates} illustrates one such example of a network in this family. While we have depicted a specific source-to-party configuration, the method works for any arrangement of sources and parties where each party receives exactly two LVs (i.e. each party receives a physical system from exactly two sources). We will focus on networks with tripartite sources, however our method also holds for $P$-partite sources (for $P\geq 3$). Furthermore, we require that no two sources signal to exactly the same set of parties\footnote{Under network-locality, separate sources are independent and thus carry different LVs. If two sources influence exactly the same parties, their effects in the observed statistics are indistinguishable, so without loss of generality, we can represent them by a single LV.}. For a ring network with $N$ parties, $A_1,\dots,A_N$, satisfying
\begin{align}
    N\geq 6, && N=0\mod 3,
\end{align}
the network has $M=2N/3$ independent sources, $S_1,\dots,S_M$. Each source $S_m$ is assigned its own LV $\lambda_m$ and sends information to exactly three parties. 

In general, there are no restrictions on the physical setup that the network displays. Here, however, we will study the behaviors $p(\outa)$ that arise from a photonic experiment similar to Ref.~\cite{abiuso2022single}. In the following, we outline the quantum description of the specific photonic experiment, where the sources emit quantum systems (rather than have an associated LV) and the parties perform quantum measurements (rather than have local response functions).

Let us start by discussing the state preparation of the system in the network. Each source $S_m$ prepare an identical single-photon W state to distribute to three parties $\{A_u,A_v,A_w\}$,
\begin{align}\label{eqn:w_state}
    \ket{\psi}_{A_uA_vA_w}=\frac{1}{\sqrt{3}}\bigl(\ket{100}+\ket{010}+\ket{001}\bigr)_{A_uA_vA_w},
\end{align}
where $\ket{\cdot}_{A_uA_vA_w}$ denotes Fock states with a photon being sent to one party and zero photons to the remaining two parties. The joint state in the network is thus
\begin{align}
    \ket{\Psi}\equiv \ket{\psi}_{A_{N-1}A_1A_2}\otimes \dots \otimes \ket{\psi}_{A_{N}A_{1}A_3},
\end{align}
where we have the tensor product of $M$ tripartite states, each corresponding to one of the sources in Fig.~\ref{fig:ring_network_wstates}. 

Now let us discuss the measurement devices each party operates in this network. Each party $X\in\{A_1,\dots,A_N\}$, receives photons on two modes $(X_1,X_2)$, which they then mix on an ideally lossless beamsplitter. The action of the beamsplitter on the creation operator vector $(\hat{a}_{X_1}^\dagger, \hat{a}_{X_2}^\dagger)^T$ is
\begin{align}
    B_{X_1X_2}=
    \begin{pmatrix}
        \sqrt{t} & -e^{-i\phi}\sqrt{1-t}\\
        e^{i\phi}\sqrt{1-t} & \sqrt{t}
    \end{pmatrix},
\end{align}
where $t$ is the transmissivity (set equal for all parties), and we set $\phi=0$. The photons are then detected by click/no-click ($\blacksquare/\square$) detectors in each mode, denoted by
\begin{align}
    D^{(\square)}=\out{0}, &&
    D^{(\blacksquare)}=\mathbb{I} - \out{0}.
\end{align}
The apparatus at each party is shown in Fig.~\ref{fig:ring_network_wstates}.
The four outcome positive operator-valued measure (POVM) on the modes $(X_1,X_2)$ can be written as follows:
\eqalign{
    \Pi^{(a_n)}_{t}=
    B_{X_1X_2}^\dagger\bigl(D^{(l)}_{X_1}\otimes D^{(r)}_{X_2}\bigr)B_{X_1X_2},\\
    a_n\in\{\xx{0},\xx{L},\xx{R},\xx{2}\},\\
    l,r\in\{\square,\blacksquare\},
}
where $a_n$ indicates the output of party $n$. The possible outputs for party $A_n$ are: no clicks ($a_n=\xx{0}$), a click in the left mode $(a_n=\xx{L})$, a click in the right mode $(a_n=\xx{R})$, and clicks in both modes $(a_n=\xx{2})$. These outcomes correspond to the following detector projectors: $(l,r)=(\square,\square), (\blacksquare,\square), (\square,\blacksquare), (\blacksquare, \blacksquare)$, respectively.

Since the detectors are non-photon-number-resolving, the number of clicks do not correspond to the number of input photons for a party. For example, two incoming photons which are detected in the same mode will only register a single click, $\xx{L}$ or $\xx{R}$.

The observed statistics in this experiment are collected via the joint distribution $p(\outa)$, which can then be written as
\begin{align}\label{eqn:quant_joint_stats}
    p(\outa)=
    \Tr\Bigl[
      \out{\Psi}\bigl(\Pi^{(a_1)}_t\otimes\dots\otimes \Pi^{(a_N)}_t\bigr)
    \Bigr].
\end{align}
We want to determine whether $p(\outa)$ admits the following network-local decomposition:
\eqalign{\label{eqn:network_local_cond}
    p(\outa) =\int \dots\int &d\lambda_1\dots d\lambda_M \\
    &\times \prod_{m=1}^M p_m(\lambda_m) \prod_{n=1}^N p_n(a_n\cond \bar{\lambda}_n),
}
where $\bar{\lambda}_n$ indicates the pair of LVs party $A_n$ receives (i.e. the LVs that come from the sources the party is connected to). For example, $\bar{\lambda}_1=(\lambda_1,\lambda_M)$ in Fig.~\ref{fig:ring_network_wstates}. Note the product of the LV distributions enforces network-locality as opposed to standard Bell locality.
Lastly, we assume photon number conservation, meaning we do not need to consider what happens in the case of null events.

\section{LP Procedure for the Ring Network}\label{sec:lp_procedure_ring_network}

Here we show how to derive the constraints in \textit{Class 3}, \textit{Class 4}, and \textit{Class 5} for the photonic experiment from Sec.~\ref{sec:ring_network_construction} associated with an $N$-party ring network. Note \textit{Class 1} and \textit{Class 2} are trivially derived. We start by showing some useful properties of LV models in this network (\textit{Step 1} and \textit{Step 2}), and then construct each class of constraints.

\subsection*{\textbf{Step 1: Enumerate LV Space}}
First we want to find a useful enumeration for the LV space $\mathcal{D}=\{\strat_j\}$, which specifies a local strategy for every observed outcome. We choose to define a strategy as follows.

\begin{definition}[Strategy]\label{def:strategy}
    For a network with $M$ single-photon sources signaling to three parties $\{A_u,A_v,A_w\}$, for $m\in M$ let $\lambda_m$ indicate which party receives a photon from source $S_m$. Hence each $\lambda_m$ can take on three values
    \begin{align}
        \lambda_m\in\{A_u,A_v,A_w\},
    \end{align}
    which we will call the target list of $\lambda_m$. The expression $\lambda_m=A_u$ means ``$\lambda_m$ tells $S_m$ to send a photon to $A_u$". Let us define a strategy as the string of all LV assignments,
    \begin{align}
        \strat_j=(\lambda_1,\dots,\lambda_m,\dots,\lambda_M), && \lambda_m \in \{A_u,A_v,A_w\},
    \end{align}
    where each target list is determined by the three parties $S_m$ is connected to. Note a party can only have non-zero outputs if it receives a photon from at least one LV.
\end{definition}

Notice here we have used the words ``sending a photon'' to explain pictorially what the local source does, but this does not require a quantum model. Indeed, the point of enumerating strategies is to describe a network-local model. For example, we could imagine sources distributing classical objects. Nevertheless, we will use this jargon as a shorthand throughout the manuscript, but we emphasize this network-local model can also be captured in purely operational terms. One could think of each $\lambda_m$ as indicating which parties are allowed to observe a click on their detectors: if the strategies are set so that a party `receives no photon' from both of its sources, then the only output this party can produce is $(\square,\square)$; if they are set so that a party `receives a photon' only from one source, then that party may output either $(\square,\blacksquare)$ or $(\blacksquare,\square)$ depending on their local response function; finally, if the strategies are set so that a party `receives a photon' from both its sources, then the party may output $(\square,\blacksquare)$, $(\blacksquare,\square)$, or $(\blacksquare,\blacksquare)$. Notice this LV model cannot produce outcomes of the form ``all parties output $(\blacksquare,\blacksquare)$'' or ``all parties output $(\square,\square)$'', which one can in principle produce trivially with more general local models. This is not a restriction of our network-local model however, as we only examine a classical model of an experiment that conserves photon number and therefore cannot produce those outcomes to begin with. As we show below, there exists a valid strategy $\strat_j\in\mathcal{D}$ for every realizable outcome $\outa_i\in\mathcal{O}$.

To fully enumerate $\mathcal{D}$, we take the Cartesian product of all the target lists of the LVs. As each of the $M$ LVs sends a photon to one of three parties, the total number of strategies is $|\mathcal{D}|= 3^M$.

To ensure this is a valid enumeration of $\mathcal{D}$, we need to verify there exists a valid strategy $\strat_j\in\mathcal{D}$ for every realizable outcome $\outa_i\in\mathcal{O}$. We now construct a map $F:\mathcal{D}\mapsto \mathcal{O}$ to ensure the enumeration supports every observed outcome. Note Def.~\ref{def:strategy} only specifies if a party $A_n$ receives a photon, and not whether it detects said photon in the $\xx{L}$ or $\xx{R}$ modes. This motivates defining a coarse-graining of the single-click outcomes, call it $\xx{X}=\{\xx{L},\xx{R}\}$. Then $a_n=\xx{X}$ means party $A_n$ detected a click in either the $\xx{L}$ mode or the $\xx{R}$ mode.
\begin{definition}[Outcome pattern]\label{def:outcome_pattern}
    For a given $\outa_i\in\mathcal{O}$, let $\bar{\outa}$ denote the single-click coarse-grained outcome, where every instance of a $\xx{L}$ or $\xx{R}$ output has been replaced with $\xx{X}$. Let us now use ``outcome pattern" as a shorthand for ``single-click coarse-grained outcome". Thus, $\bar{\outa}$ represents the set of outcome strings compatible with a given outcome pattern. Let
    \begin{align}
        \bar{n}=\{n:a_n=\xx{L}\text{ or }a_n=\xx{R}\},&& a_n\in\bar{\outa},
    \end{align}
    be the set of party indices with value $a_n=\xx{X}$. Then the marginal probability of $\bar{\outa}$ is
    \begin{align}
        p(\bar{\outa})=\sum_{\{a_n\in\{\xx{L},\xx{R}\}\}_{n\in\bar{n}}} p(a_1,\dots,a_n,\dots,a_N),
    \end{align}
    where $n\in\bar{n}$ indicates which $a_n$ appear in the nested sums.
\end{definition}

We reconstruct the image of a given strategy $\strat_j$ under $F$ using the following procedure:
\begin{enumerate}
    \item Initialize an empty outcome string $\outa$.
    \item \textbf{Replacement:} If $\lambda_m=A_n$, set $a_n\in\outa$ to $\xx{X}$ (thus including $a_n=\xx{L}$ and $a_n=\xx{R}$ outcomes in the image).
    \item Apply the replacement rule for all $\lambda_m\in\strat_j$ and include the resulting outcome $\outa$ in the image.
    \item \textbf{Collision:} If two LVs send a photon to the same party $\lambda_m=\lambda_{m'}=A_n$, include the outcomes $\outa$ for which $a_n=\xx{2}$ and $a_n=\xx{X}$ in the image.
\end{enumerate}
The collision rule is a consequence of the non-photon-number-resolving detectors. As $\lambda_m$ does not specify the mode in which the photon is detected, we must include all the fine-grainings of $\xx{X}$ outcomes. If the number of single-click outputs in a pattern is $\#\xx{X}=k$, the expanded set will have $2^k$ outcomes.

Hence, the full \textit{Step 1} proceeds as follows: we enumerate strategies according to Def.~\ref{def:strategy} and construct the image of $\mathcal{D}$ under $F$ as defined in the above procedure. We then compare it against the true observed set of outcomes $\mathcal{O}$. If they are equivalent, we have verified our enumeration produced at least one strategy in the support of every realizable outcome, and hence is valid.

\subsection*{\textbf{Step 2: Isolate an Amenable Outcome Subset}}
This step is a necessary precursor for the domain asymmetry constraints in \textit{Class 3}. In particular, we want to isolate outcomes whose pre-image can be decomposed into disjoint regions:
\begin{align}\label{eqn:key_property_op}
    \mathcal{O}_p:\mathcal{S}_p=\mathcal{S}_p^{(1)}\sqcup\mathcal{S}_p^{(2)}, && \mathcal{O}_p\subset\mathcal{O}_S,\: \mathcal{S}_p\subset \mathcal{S}.
\end{align}
In other words, we are looking for events which can happen in two unique ways. This will help us build towards an analytical form for the domain asymmetries $\Delta_{\mathcal{O}_p}$. Let us first define our choice of $\mathcal{O}_S\subset\mathcal{O}$ and its pre-image $\mathcal{S}\subset\mathcal{D}$, over which we define the joint distribution $q(\outa,\strat)$.

\begin{definition}[Outcome Subset]\label{def:o_s}
    Let $\mathcal{O}_S\subset\mathcal{O}$ be the subset of outcomes satisfying
    \begin{align}
        \mathcal{O}_S=\{\outa:\# \xx{X}=M \text{ and }\#\xx{0}=N-M\},
    \end{align}
    where the number of single-click outputs is equivalent to the total number of photons $M$, and the number of zero-click outputs is equivalent to $N-M$.
\end{definition}

Due to photon number conservation, every photon is explicitly accounted for as a click in the outcome, and hence no party can receive photons from the two sources simultaneously. We have chosen this specific $\mathcal{O}_S$ as it gives rise to two operationally useful properties. The first relates to strategies in its pre-image, $\mathcal{S}=F^{-1}(\mathcal{O}_S)$, and the second relates to strategies not in its pre-image.

\begin{property}[Condition on Strategies in $\mathcal{S}$]\label{prop:S_rules}
    $\mathcal{S}$ cannot contain strategies where two $\lambda_m$ send a photon to the same party, $\lambda_m= \lambda_{m'}=A_n$. Alternatively, we can define $\mathcal{S}$ as,
    \begin{align}
        \mathcal{S}=\{\strat:\#A_n\leq 1\},&& \forall n\in 1,\dots, N,
    \end{align}
    meaning each $A_n$ can appear at most once in $\strat_j$.
\end{property}

\noindent\textit{Proof:} Consider a strategy with $\lambda_m=\lambda_{m'}=A_n$. By Def.~\ref{def:strategy}, each $\lambda_m$ must send exactly one photon to one of three parties it is connected to. If both photons sent by $\lambda_m$ and $\lambda_{m'}$ go into the same detector, the output at that party will be $a_n=\xx{X}$. However, this means the total number of single-click outputs in the outcome string will be less than the total number of photons, $\#\xx{X}< M$, meaning it cannot be in the set $\mathcal{O}_S$. If the photons sent by $\lambda_m$ and $\lambda_{m'}$ end up in different detectors, the output at that party will be $a_n=\xx{2}$. However, outcomes in $\mathcal{O}_S$ cannot contain $\xx{2}$ outputs, by Def.~\ref{def:o_s}. Hence, none of the outcomes in the image of a strategy with $\lambda_m=\lambda_{m'}=A_n$ can be in $\mathcal{O}_S$. \hfill $\blacksquare$

\begin{property}[Condition on Strategies not in $\mathcal{S}$]\label{prop:S_supports_only_Os}
    Every outcome not in the subset $\mathcal{O}_S$, $\outa\in\mathcal{O}\setminus\mathcal{O}_S$, must have at least one instance of $\lambda_m=\lambda_{m'}$ in its corresponding strategy. Hence
    \begin{align}
        F(\mathcal{S})&\mapsto \mathcal{O}_S,
    \end{align}
    meaning $\mathcal{S}$ must contain strategies only in the support of outcomes in $\mathcal{O}_S$.
\end{property}

\noindent\textit{Proof:} Suppose there exists a strategy $\strat\in\mathcal{S}$ such that $F(\strat)\mapsto \outa\in\mathcal{O}\setminus\mathcal{O}_S$. Note that every outcome $\outa\in\mathcal{O}\setminus\mathcal{O}_S$ must satisfy
\begin{align}
    \mathcal{O}\setminus\mathcal{O}_S=\{\outa:\#\xx{X}<M\text{ and }\#\xx{0}>N-M\}.
\end{align}
Under photon number conservation, the only way for $\#\xx{X}<M$ is if at least two photons are sent to the same party. But if the strategy is in the support of $\mathcal{O}_S$, it must mean there are no instances of $\lambda_m=\lambda_{m'}$, as shown in Prop.~\ref{prop:S_rules}. We have a contradiction. Thus, there are no strategies in $\mathcal{S}$ which map to outcomes outside of $\mathcal{O}_S$. \hfill $\blacksquare$

\bigskip

With these properties established, we can now search for events in $\mathcal{O}_S$ which correspond to disjoint subsets in LV space. We want to compute the probability of each subset in LV space from observed statistics, but this is difficult, as we usually cannot access information about the values of the LVs from $p(\outa)$ alone. However, the fact that our network has fewer photons than number of parties allows us to isolate the value of a single $\lambda_m$ from the observed statistics. To demonstrate why this is true, let us first establish a few definitions.

\begin{definition}[Outcome Substring]\label{def:outcome_substring}
    Consider the three parties in the target list of $\lambda_m$, $\{A_u,A_v,A_w\}$. Suppose one party registers a single-click output, while the remaining two parties register a zero-click output. Denote this three-party outcome substring as
    \begin{align}
        \outa^{(m,n)}=
        \begin{cases}
            a_n=\xx{X}, & n\in\{u,v,w\}\\
            (a_u,a_v,a_w)\setminus a_n=\xx{0},
        \end{cases}
    \end{align}
    where $m$ fixes the target list determined by $\lambda_m$, and $n$ indicates which $A_n$ in the target list detected a single-click.
\end{definition}

Given an outcome $\outa\in\mathcal{O}$, we say it is compatible with a substring $\outa^{(m,n)}$ if the three outcomes $(a_u,a_v,a_w)\in\outa$ match the conditions in Def.~\ref{def:outcome_substring}. Here $\xx{X}$ indicates $a_n=\xx{L}$ and $a_n=\xx{R}$ outputs are both compatible. If $\outa\in\mathcal{O}$ is compatible with $\outa^{(m,n)}$ we denote it $\outa^{(m,n)}\in\outa$. 

Let us now define the set of outcome patterns in $\mathcal{O}_S$ compatible with a given substring. For future definitions, we will denote two sets: one that corresponds to $\outa^{(m,n)}$ and one that corresponds to $\outa^{(m',n)}$, as follows:
\begin{align}
    \mathcal{O}_p^{(1)}&=\{\bar{\outa}\in\mathcal{O}_S: \outa^{(m,n)}\in\bar{\outa}\} \label{eqn:op1},\\
    \mathcal{O}_p^{(2)}&=\{\bar{\outa}\in\mathcal{O}_S: \outa^{(m',n)}\in\bar{\outa}\} \label{eqn:op2}.
\end{align}

\begin{definition}[Remaining LVs]\label{def:strat_rest}
    Let us denote the set of LVs that are not $\lambda_m$ as follows:
    \begin{align}
        \stratrest{m}&=\strat\setminus \lambda_m=(\lambda_1,\dots,\lambda_{m-1},\lambda_{m+1},\dots,\lambda_M).
    \end{align}
\end{definition}

Def.~\ref{def:outcome_substring} alone is not enough information to make inferences about the values of LVs. However, if we restrict our attention to outcomes $\outa\in\mathcal{O}_S$ compatible with a given $\outa^{(m,n)}$, we know no two LVs can send a photon to the same party. We also know the remaining two parties in the target list of $\lambda_m$ must not receive a photon as they output $\xx{0}$. Thus, the set of LVs $\stratrest{m}$ must not be interacting with any of the parties $\lambda_m$ is connected to. This allows us to derive conditions on the strategies in the support of outcomes in $\mathcal{O}_S$ compatible with a given substring $\outa^{(m,n)}$.

\begin{theorem}[LV Value]\label{thm:lv_value}
    Consider outcomes $\outa\in\mathcal{O}_S$ with $a_n=\xx{X}$. Suppose the two LVs $A_n$ receives have target lists
    \begin{align}
        \lambda_m\in\{A_u,A_v,A_w\},\qquad \lambda_{m'}\in\{A_{u'},A_{v'},A_{w'}\},\\
        \text{with }\{A_u,A_v,A_w\}\neq\{A_{u'},A_{v'},A_{w'}\},\\
        \text{and }A_w= A_{w'}=A_n.
    \end{align}
    Both lists contain $A_n$, and for sources distributing to $P$ parties, their maximum overlap can extend to $P-1$ parties. Now consider the joint target list minus $A_n$, $\{A_u,A_v,A_{u'},A_{v'}\}$. If we observe the outcomes in this joint target list are compatible with exactly one of two outcome substrings, then we can determine which LV $(\lambda_m,\lambda_{m'})$ sent $A_n$ a photon. In particular,
    \begin{align}
        (a_u,a_v)=\xx{0} &\implies \lambda_m=A_n, \label{eqn:thm1_claim1}\\
        (a_{u'},a_{v'})=\xx{0} &\implies \lambda_{m'}=A_n.\label{eqn:thm1_claim2}
    \end{align}
    In other words, the outcomes in $\mathcal{O}_S$ compatible with the substring $\outa^{(m,n)}$ imply $\lambda_m=A_n$ and those compatible with the substring $\outa^{(m',n)}$ imply $\lambda_{m'}=A_n$.
\end{theorem}
    
Thm.~\ref{thm:lv_value} is the key insight that allows us to access information about the value of a specific $\lambda_m$ from the observed statistics alone. Intuitively, because there are fewer photons than number of parties, we can analyze the positions of the $\xx{0}$ outputs to isolate outcomes where $\stratrest{m}$ does not interact with any party that $\lambda_m$ can interact with. This also underlies our definition of $\mathcal{O}_S$, as it excludes strategies where multiple LVs send a photon to the same party. Consequently, we can infer $\lambda_m$ must have sent a photon to $A_n$ if an outcome $\outa\in\mathcal{O}_S$ is compatible with the substring $\outa^{(m,n)}$.\footnote{For example, consider the outcome substrings $\outa^{(1,1)}$ and $\outa^{(M,1)}$ associated with the network in Fig.~\ref{fig:ring_network_wstates}. We have $\lambda_1\in\{A_{N-1},A_1,A_2\}$ and $\lambda_M\in\{A_N,A_1,A_3\}$. The remaining joint target list is $\{A_{N-1},A_N,A_2,A_3\}$. Thus, $a_{N-1}$, $a_N$, $a_2$, and $a_3$ are not allowed to be simultaneously $\xx{0}$. Now, $\outa^{(M,1)}$ includes all outcomes with $a_N=a_3=\xx{0}$, and hence have either $a_{N-1}\neq \xx{0}$ or $a_2\neq \xx{0}$. Therefore, these outcomes are not compatible with the substring $\outa^{(1,1)}$. Similarly, outcomes that yield $\outa^{(1,1)}$ have either $a_N\neq \xx{0}$ or $a_3\neq \xx{0}$, and hence are not compatible with the substring $\outa^{(M,1)}$.} For the full proof of Thm.~\ref{thm:lv_value}, see Appendix~\ref{appendix:proof_lv_value_thm}.

The quantity of interest we compute from observed statistics and later relate to the weight of subsets in LV space is the marginal probability $p(\outa\in\mathcal{O}_p^{(1)})$ for different outcome substrings $\outa^{(m,n)}$. Recall the domain asymmetry from Eqn.~\eqref{eqn:general_domain_constraint} is defined in terms of joint probabilities $q(\outa,\strat)$ on the subsets $\mathcal{O}_S$ and $\mathcal{S}$. Thus we write $q(\outa^{(m,n)})$ to mean $p(\outa\in\mathcal{O}_p^{(1)})$ renormalized over $\mathcal{O}_S$. By linearity, we have
\begin{align}
    &q\left(\outa^{(m,n)}\right)\equiv q\left(\outa\in\mathcal{O}_p^{(1)}\right),\\
    &=\sum_{\bar{\outa}\in\mathcal{O}_p^{(1)}}\left(\sum_{\{a_n\in\{\xx{L},\xx{R}\}\}_{n\in\bar{n}}}q(\outa)\right),\label{eqn:thm2_line1}\\
    &=\sum_{\bar{\outa}\in\mathcal{O}_p^{(1)}}\left(\sum_{\{a_n\in\{\xx{L},\xx{R}\}\}_{n\in\bar{n}}}\frac{p(\outa)}{\sum_{\outa'\in\mathcal{O}_S}p(\outa')}\right), \label{eqn:prob_amn}
\end{align}
where we have used Def.~\ref{def:outcome_pattern} in the first line. Note the set of indices $\bar{n}$ defining the nested sums in the parenthetical is determined by the outcome pattern $\bar{\outa}$ in the outermost sum. We write Eqn.~\eqref{eqn:prob_amn} explicitly in terms of $p(\outa)$, and note it is a scalar value. A similar statement holds for $q(\outa^{(m',n)})$.

\begin{theorem}[Probability of an LV Subset]\label{thm:prob_lv_subset}
    Given a subset of outcome patterns $\mathcal{O}_p^{(1)}$ or $\mathcal{O}_p^{(2)}$, the conditions on the strategies in its pre-image must satisfy the following, respectively:
    \eqalign{\label{eqn:sp1}
        \mathcal{S}_p^{(1)}=\{\strat&\in\mathcal{S}:\\
        &\lambda_m=A_n,\stratrest{m}\neq \{A_u,A_v,A_w\}\},
    }
    \eqalign{\label{eqn:sp2}
        \mathcal{S}_p^{(2)}=\{\strat&\in\mathcal{S}:\\
        &\lambda_{m'}=A_n,\stratrest{m'}\neq \{A_u',A_v',A_w'\}\}.
    }
    Furthermore, the total probability weight of $\mathcal{S}_p^{(1)}$ and $\mathcal{O}_p^{(1)}$ given by $\mu$ are equivalent:
    \begin{align}
        \mu\left(\strat\in\mathcal{S}_p^{(1)}\right) =\mu\left(\outa\in\mathcal{O}_p^{(1)}\right).\label{eqn:weight_sp1}
    \end{align}
    We can show $\mu(\outa\in\mathcal{O}_S)=\mu(\strat\in\mathcal{S})$ and hence,
    \begin{align}
        q\left(\strat\in\mathcal{S}_p^{(1)}\right)=q\left(\outa\in\mathcal{O}_p^{(1)}\right).\label{eqn:weight_sp1_q}
    \end{align}
    A similar statement holds for $\mathcal{S}_p^{(2)}$ and $\mathcal{O}_p^{(2)}$.
\end{theorem}
Thm.~\ref{thm:prob_lv_subset} allows us to quantify the weight of a subset $\mathcal{S}_p^{(1)}$ in LV space via Eqn.~\eqref{eqn:prob_amn}, using the fact that $\mu(\outa)$ can be computed directly from $p(\outa)$. This relationship holds for any network in our family of ring networks. For the full proof of Thm.~\ref{thm:prob_lv_subset}, see Appendix~\ref{appendix:proof_thm2}.

We now have two events in $\mathcal{O}_S$ ($a_n=\xx{L}$ or $a_n=\xx{R}$) which can happen in two unique ways (either in a way compatible with $\outa^{(m,n)}$ or in a way compatible with $\outa^{(m',n)}$). Note these are not the only ways the event $a_n=\xx{X}$ in $\mathcal{O}_S$ can happen.\footnote{For example, consider the 6-party, 4-source network in Fig.~\ref{fig:6p4s_ring_network}. The outcome $\bar{\outa}=(\xx{X},\xx{X},\xx{X},\xx{0},\xx{0},\xx{X})$ has $a_1=\xx{X}$ but $\bar{\outa}$ is not compatible with $\outa^{(1,1)}$ (since $a_2=a_2=\xx{X}$) nor $\outa^{(4,1)}$ (since $a_6=\xx{X}$). This outcome pattern can arise via the global strategy $\strat=(\lambda_1=A_1,\lambda_2=A_2,\lambda_3=A_3,\lambda_4=A_6)$.} Rather, the key idea is to artificially construct two mutually exclusive ways which $a_n=\xx{X}$ can occur.

We define the subset $\mathcal{O}_p$ as the union of outcome patterns compatible with either the substring $\outa^{(m,n)}$ or $\outa^{(m',n)}$ in $\mathcal{O}_S$:
\begin{align}\label{eq:theop}
\mathcal{O}_p&=\mathcal{O}_p^{(1)}\cup\mathcal{O}_p^{(2)},
\end{align}
where $\mathcal{O}_p^{(1)}$ and $\mathcal{O}_p^{(2)}$ are defined as in Eqns.~\eqref{eqn:op1} and \eqref{eqn:op2}. The pre-image of this subset can be decomposed as
\begin{align}\label{eqn:thesp}
    \mathcal{S}_p&=\mathcal{S}_p^{(1)}\sqcup \mathcal{S}_p^{(2)},
\end{align}
by Thm.~\ref{thm:prob_lv_subset}, where $\mathcal{S}_p^{(1)}$ and $\mathcal{S}_p^{(2)}$ are defined as in Eqns.~\eqref{eqn:sp1} and \eqref{eqn:sp2}. By definition, $\mathcal{S}_p^{(1)}$ and $\mathcal{S}_p^{(2)}$ are disjoint\footnote{In fact, $\mathcal{O}_p^{(1)}$ and $\mathcal{O}_p^{(2)}$ are also disjoint, but in principle, they need not be. The condition for $\mathcal{O}_p$ given in Eqn.~\eqref{eqn:key_property_op} is a sufficient one.}. With these definitions, let us define one more property.

\begin{property}[$F(\mathcal{S}_p)=\mathcal{O}_p$]\label{prop:f_sp_op}
    For every outcome not in the subset $\mathcal{O}_p$, $\outa\in\mathcal{O}_S\setminus\mathcal{O}_p$, the joint probability satisfies
    \begin{align}
        \mu(\outa,\strat)=0, && \forall \strat\in\mathcal{S}_p,\\
        F(\mathcal{S}_p)=\mathcal{O}_p.
    \end{align}
    In other words, strategies in $\mathcal{S}_p$ do not produce outcomes outside $\mathcal{O}_p$.
\end{property}
\noindent\textit{Proof:} Suppose there exists a strategy $\strat\in\mathcal{S}_p$ such that $F(\strat)\mapsto \outa\in\mathcal{O}_S\setminus\mathcal{O}_p$. By Thm.~\ref{thm:prob_lv_subset}, either $\lambda_m=A_n$ or $\lambda_{m'}=A_n$, meaning $a_n=\xx{X}$. Note outcomes $\outa\in\mathcal{O}_S\setminus\mathcal{O}_p$ cannot be compatible with either $\outa^{(m,n)}$ nor $\outa^{(m',n)}$. Hence one of the parties in the remaining target list $(a_u,a_v,a_w,a_{u'},a_{v'},a_{w'})\setminus a_n\neq \xx{0}$. However, this is only possible if $\stratrest{m}\in\{A_u,A_v,A_w\}$ or $\stratrest{m'}\in\{A_{u'},A_{v'},A_{w'}\}$. This contradicts the definition of $\mathcal{S}_p$. Thus, there are no strategies in $\mathcal{S}_p$ which map to outcomes outside of $\mathcal{O}_p$. \hfill $\blacksquare$

\subsection*{\textbf{Step 3a: Construct LP Constraints Class 3}}
We now construct the \textit{Class 3} constraints for the ring network. Recall the network-locality condition requires the distribution of the strategies to factorize as
\begin{align}\label{eqn:p_lv_independence}
    \mu(\strat_j)=\mu(\lambda_1,\dots,\lambda_M)=\prod_{m=1}^M \mu_m(\lambda_m).
\end{align}
We can write Eqn.~\eqref{eqn:sp1} in terms of the LV values in $\mathcal{S}_p^{(1)}$ as follows:
\begin{align}
    \mu\left(\outa\in\mathcal{O}_p^{(1)}\right)&=\mu\big(\lambda_m=A_n,\stratrest{m}\neq\{A_u,A_v,A_w\}\big), \\
    \mu\left(\lambda_m=A_n\right)&=\frac{\mu\left(\outa\in\mathcal{O}_p^{(1)}\right)}{\mu\left(\stratrest{m}\neq\{A_u,A_v,A_w\}\right)},\label{eq:64}
\end{align}
where we have used Eqn.~\eqref{eqn:p_lv_independence} in the second line. Since we consider $\strat\in\mathcal{S}_p^{(1)}$, the set of outcomes compatible with $\stratrest{m}\neq\{A_u,A_v,A_w\}$ can be  partitioned into three disjoint sets of outcome patterns corresponding to $\outa^{(m,u)}$, $\outa^{(m,v)}$, or $\outa^{(m,w)}$. Let us denote these sets by $\mathcal{O}_{u}^{(1)}$, $\mathcal{O}_{v}^{(1)}$, and $\mathcal{O}_{w}^{(1)}$, respectively. Under the assumption that there exists an LV model for our observed outcome statistics, then one can leverage Eqn.~\eqref{eqn:weight_sp1} to  rewrite the denominator in Eqn.~\eqref{eq:64} as
\begin{align}
    \mu\left(\strat^{(\text{rest})}\neq \{A_u,A_v,A_w\}\right)= \sum_{n\in\{u,v,w\}} \mu\left(\outa\in\mathcal{O}_n^{(1)}\right).
\end{align}
Since $\mu(\outa)$ can be computed via the observed statistics $p(\outa)$, $\mu(\lambda_m=A_n)$ can be written in terms of $p(\outa)$ as
\begin{align}\label{eqn:prob_lm_an}
    \mu(\lambda_m=A_n)=\frac{p(\outa\in\mathcal{O}_{n}^{(1)})}{p(\outa\in\mathcal{O}_{u}^{(1)})+p(\outa\in\mathcal{O}_{v}^{(1)})+p(\outa\in\mathcal{O}_{w}^{(1)})}.
\end{align}
When our observed statistics admits a network-local model, we can compute each marginal probability $p(\outa\in\mathcal{O}_{n}^{(1)})$ and relate them to the probability of each $\mu(\lambda_m=A_n)$. Using Eqn.~\eqref{eqn:p_lv_independence}, we can derive the probability of every strategy $\mu(\strat_j)$ as the product of each of the LV assignments, $\mu_m(\lambda_m)$, given by Eqn.~\eqref{eqn:prob_lm_an}.  Additionally, once we know $\mu(\strat)$, we can relate it to the distribution $q(\strat)$ via Eqn.~\eqref{eqn:q_lambda}.

\bigskip

\begin{sumblocknt}{}
 \textbf{\textit{Class 3} constraints for ring network.} \\
 For each $\lambda_m \in \{A_u,A_v,A_w\}$, compute $\mu(\lambda_m=A_n)$ from $p(\outa)$ via Eqn.~\eqref{eqn:prob_lm_an}. 
 
 Then, for each $\lambda_j \in \mathcal{D}$, compute $\mu(\strat_j)$ from $p(\outa_j)$ via Eqn.~\eqref{eqn:p_lv_independence}. 

 Finally, for $\strat \in \mathcal{S}$ \textbf{impose} constraints on $q(\strat)$ from $\mu(\strat)$ via Eqn.~\eqref{eqn:q_lambda}.  
\end{sumblocknt}
 
\subsection*{\textbf{Step 3b: Construct LP Constraints Class 4}}
We now construct the \textit{Class 4} constraints. Note that when the observed statistics $p(\outa)$ admit a network-local model, the statistics a single-party observes locally (given by the marginal $p_n(a)$ in Eqn.~\eqref{eqn:spm}) must only depend on the values of the LVs of the sources that party is connected to.

In the case of the ring network we consider, each party $A_n$ is connected to two sources ($S_m$ and $S_{m'}$) whose LVs are denoted by $\lambda_m$ and $\lambda_{m'}$. The idea is that if we have two strategies 
\begin{align}
     \strat&\equiv (\lambda_1,\dots,\lambda_m,\dots,\lambda_{m'},\dots,\lambda_M),\\
    \strat'&\equiv (\lambda_1',\dots,\lambda_m',\dots,\lambda_{m'}',\dots,\lambda_M'),
\end{align}
such that $\lambda_m=\lambda'_m$ and $\lambda_{m'}=\lambda'_{m'}$ then 
\begin{align}\label{eqn:p_cond_ind}
    \mu_n(a\cond\strat)=\mu_n(a\cond\strat')\,.
\end{align}
To see how Eqn,~\eqref{eqn:p_cond_ind} translates into constraints for $q(\outa,\strat)$, begin by noticing these constraints correspond to 
\begin{align}
    \frac{1}{\mu(\strat)} \sum_{\outa \in \mathcal{O} \,|\, a_n = a} \mu(\outa,\strat)= \frac{1}{\mu(\strat')} \sum_{\outa \in \mathcal{O} \,|\, a_n = a}  \mu(\outa,\strat')\,.
\end{align}
Now, if we focus on $\strat,\strat' \in \mathcal{S}$, then:
\begin{align}
    \frac{1}{\mu(\strat)} \sum_{\outa \in \mathcal{O}_S \,|\, a_n = a} \mu(\outa,\strat)= \frac{1}{\mu(\strat')} \sum_{\outa \in \mathcal{O}_S \,|\, a_n = a}  \mu(\outa,\strat')\,,
\end{align}
since $\mu(\outa,\strat) = 0$ whenever $\outa \in \mathcal{O} \setminus \mathcal{O}_S$ and $\strat \in \mathcal{S}$. Dividing across by the renormalization over $\mathcal{S}$, $\sum_{\strat'\in\mathcal{S}}\mu(\strat')$, and using Eqn.~\eqref{eqn:q_lambda}, we have
\begin{align} \label{eq:themarg}
    \frac{1}{\mu(\strat)} \sum_{\outa \in \mathcal{O}_S \,|\, a_n = a} q(\outa,\strat)= \frac{1}{\mu(\strat')} \sum_{\outa \in \mathcal{O}_S \,|\, a_n = a}  q(\outa,\strat')\,.
\end{align}
Now, from Eqn.~\eqref{eqn:prob_lm_an} we can compute $\mu(\strat)$ from the observed data $p(\outa)$. With these, we can construct a linear constraint on $q(\outa,\strat)$ from Eqn.~\eqref{eq:themarg} as follows.

\bigskip

\begin{sumblocknt}{}
  \textbf{\textit{Class 4} constraints for ring network.} \\

 For each $n \in \{1,\ldots,N\} $, denote by $S_m$ and $S_{m'}$ the two sources party $A_n$ is connected to. For each pair of valid local variables $(\tilde{\lambda}_m,\tilde{\lambda}_{m'})$, identify all pairs of strategies $\strat,\strat' \in \mathcal{S}$ such that
 \begin{align*}
     \strat&\equiv (\lambda_1,\dots,\lambda_m,\dots,\lambda_{m'},\dots,\lambda_M),\\
    \strat'&\equiv (\lambda_1',\dots,\lambda_m',\dots,\lambda_{m'}',\dots,\lambda_M'),
 \end{align*}
 with $\lambda_m=\lambda_m'=\tilde{\lambda}_m$ and $\lambda_{m'}=\lambda_{m'}'=\tilde{\lambda}_{m'}$.

 For each such pair $(\strat,\strat')$ \textbf{impose} that 
 \begin{align*} 
    \sum_{\outa \in \mathcal{O}_S \,|\, a_n = \xx{X}} q(\outa,\strat)= \frac{\mu(\strat)}{\mu(\strat')} \sum_{\outa \in \mathcal{O}_S \,|\, a_n = \xx{X}}  q(\outa,\strat')\,.
\end{align*}
 \end{sumblocknt}

\subsection*{\textbf{Step 3c: Construct LP Constraints Class 5}} 
We now discuss how to explicitly write the constraints for \textit{Class 5} for the ring network under consideration. Let us begin by defining a new quantity (denoted by $\Gamma_{\mathcal{O}_p}$) to be computed from our observed statistics  $\{p(\outa)\}_{\outa \in \mathcal{O}}$.
For each of the $2N$ sets $\mathcal{O}_p$ defined in Eqn.~\eqref{eq:theop}, define 
\eqalign{\label{eq:dac1}
     \Gamma_{\mathcal{O}_p}=&\sum_{\bar{\outa}\in\mathcal{O}_p^{(1)}}\left(\sum_{\{a_n\in\{\xx{L},\xx{R}\}\}_{n\in\bar{n}}}\frac{p(\outa)}{\sum_{\outa'\in\mathcal{O}_S}p(\outa')}\right)  \\
     & - \sum_{\bar{\outa}\in\mathcal{O}_p^{(2)}}\left(\sum_{\{a_n\in\{\xx{L},\xx{R}\}\}_{n\in\bar{n}}}\frac{p(\outa)}{\sum_{\outa'\in\mathcal{O}_S}p(\outa')}\right)
}
From the definition of $q(\outa,\strat)$ in Eqn.~\eqref{eqn:constraints2} one obtains Eqn.~\eqref{eqn:prob_amn}, from which it follows that
\begin{align}\label{eq:dac2}
    \Gamma_{\mathcal{O}_p}=q\left(\outa^{(m,n)}\right)-q\left(\outa^{(m',n)}\right).
\end{align}
Now, by definition in Eqn.~\eqref{eqn:qa_marginal} we have,
\begin{align}
    q\left(\outa^{(m,n)}\right) = \sum_{\outa\in\mathcal{O}_p^{(1)}} q(\outa) = \sum_{\outa\in\mathcal{O}_p^{(1)}} \sum_{\strat \in \mathcal{S}} q(\outa,\strat).
\end{align}
From Eqn.~\eqref{eqn:sp1} and Prop.~\ref{prop:f_sp_op} then
\begin{align}
    q\left(\outa^{(m,n)}\right) = \sum_{\outa\in\mathcal{O}_p^{(1)}} q(\outa) = \sum_{\outa\in\mathcal{O}_p^{(1)}} \sum_{\strat \in \mathcal{S}_p^{(1)}} q(\outa,\strat).
\end{align}
It follows then that under the assumption of the existence of an LV model for the observed statistics, 
\begin{align}\label{eqn:ring_network_domain_asymmetry}
    \Delta_{\mathcal{O}_p}=q\left(\outa^{(m,n)}\right)-q\left(\outa^{(m',n)}\right),
\end{align}
where $\Delta_{\mathcal{O}_{p}} $ is defined in Eqn.~\eqref{eqn:general_domain_constraint}. Therefore
\begin{align}
    \Gamma_{\mathcal{O}_{p}}  \equiv \Delta_{\mathcal{O}_{p}}\,.
\end{align}
This gives us a systematic way to specify the domain asymmetry constraints for the ring network. 

\bigskip

\begin{sumblocknt}{}
 \textbf{\textit{Class 5} constraints for ring network.} \\
 For each of the $2N$ sets $\mathcal{O}_p$ defined in Eqn.~\eqref{eq:theop}, compute the quantities $\Delta_{\mathcal{O}_{p}}$ from the observed statistics  $\{p(\outa)\}_{\outa \in \mathcal{O}}$ via 
 \eqalign{\label{eq:58}
     \Gamma_{\mathcal{O}_p}=&\sum_{\bar{\outa}\in\mathcal{O}_p^{(1)}}\left(\sum_{\{a_n\in\{\xx{L},\xx{R}\}\}_{n\in\bar{n}}}\frac{p(\outa)}{\sum_{\outa'\in\mathcal{O}_S}p(\outa')}\right)  \\
     & - \sum_{\bar{\outa}\in\mathcal{O}_p^{(2)}}\left(\sum_{\{a_n\in\{\xx{L},\xx{R}\}\}_{n\in\bar{n}}}\frac{p(\outa)}{\sum_{\outa'\in\mathcal{O}_S}p(\outa')}\right)
}
 Now \textbf{impose} that these quantities relate to $q(\outa,\strat)$ via Eqn.~\eqref{eqn:general_domain_constraint}:
\begin{align}\label{eqn:ring_network_domain_asymmetry_constraints}
    \Delta_{\mathcal{O}_{p}}=\sum_{\outa\in\mathcal{O}_{p}}\left(\sum_{\strat\in\mathcal{S}_{p}^{(1)}} q(\outa,\strat) - \!\!\! \sum_{\strat'\in\mathcal{S}_{p}^{(2)}} q(\outa,\strat') \right).
\end{align}
\end{sumblocknt}

\section{Special Case: 6 Party 4 Source Network}\label{sec:6p4s_network}
\begin{figure}
    \centering
    \includegraphics[width=0.5\linewidth]{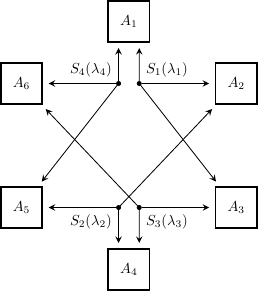}
    \caption{6-party ring network with 4 sources, characterized by their LVs, $\lambda_1$, $\lambda_2$, $\lambda_3$, and $\lambda_4$, distributing 3-party W states. Each party is equipped with a tunable beamsplitter and two detectors.}
    \label{fig:6p4s_ring_network}
\end{figure}

We now present the results for a 6-party, 4-source ring network, as shown in Fig.~\ref{fig:6p4s_ring_network}. The states the sources distribute, photonic apparatus at each party, and measurements are the same as outlined in Section~\ref{sec:ring_network_construction}. We compute the joint statistics from Eqn.~\eqref{eqn:quant_joint_stats} for the set of outcomes which satisfy photon conservation. Notice we use a slightly different distribution of sources\footnote{If we take Fig.~\eqref{fig:ring_network_wstates} for 6 parties, then parties $A_3$ and $A_6$ share two sources ($S_4$ and $S_2$), whereas in Fig.~\eqref{fig:6p4s_ring_network} those parties have only one source in common ($S_3$). } than in Fig.~\eqref{fig:ring_network_wstates}. This highlights how our method can be applied to ring networks with arbitrary source-to-party configurations.

\subsection*{Step 1: Enumerate LV Space}
We first enumerate the set of local strategies as given in Def.~\ref{def:strategy} and find $|\mathcal{D}|=3^4=81$ strategies. We determine the outcome image under $F$ for each strategy using the procedure outlined in Sec.~\ref{sec:lp_procedure_ring_network} and verify it does indeed match the set of realizable outcomes in the network, $\mathcal{O}$. Second, we isolate the outcome subset $\mathcal{O}_S$, which corresponds to 4 single-click outputs and 2 zero-click outputs, and its pre-image $\mathcal{S}$. For the single-click coarse-graining $\xx{X}=\{\xx{L},\xx{R}\}$, we have $\binom{6}{2}=15$ outcome patterns in $\mathcal{O}_S$\footnote{For the 6-party, 4-source scenario, this is equivalent to all the combinations of two $\xx{0}$ outputs. However, in general $|\mathcal{O}_S|\neq \binom{N}{N-M}$, as photon number conservation means all 3 parties in an LV's target list cannot be simultaneously $\xx{0}$.}. There are $2^{\#\xx{X}}=2^4$ possible ways to fine-grain each of these 15 patterns, which gives us $|\mathcal{O}_S|=240$ realizable outputs in our network. From $F$, we find 2 strategies in the support of every outcome pattern, giving us $|\mathcal{S}|=30$ strategies. The strategies in $\mathcal{S}$ and their corresponding outcome patterns in $\mathcal{O}_S$ are shown in Table~\ref{table:6p4s_ring_lv_strats}, where we have used the following shorthand to indicate a specific detection pattern:
\eqalign{
    \outa &= \xx{00XXXX}\\
    &\equiv(a_1=a_2=\xx{0},a_3=a_4=a_5=a_6=\xx{X}),
}
where $\xx{X}$ indicates the single-click coarse graining as usual. Let us index the strategies as $\strat_j$ for $j\in[0,29]$.

\subsection*{Step 2: Isolate Disjoint Regions of LV Space}
We define $2N$ sets $\mathcal{O}_p=\mathcal{O}_{a_n=\xx{X}}$, as in Eqn.~\eqref{eq:theop}, and isolate their corresponding pre-images $\mathcal{S}_{a_n=\xx{X}}$, as in Eqn.~\eqref{eqn:thesp}.

\subsection*{Step 3a: Constraint Class 3}
From Eqn.~\eqref{eqn:prob_lm_an}, we analytically find the distribution of strategies should be uniform: 
\begin{align}
    \mu(\lambda_m=A_n)&=1/3, && \forall m\in M,\: n\in N,\\
    \mu(\strat_j)&=1/81.
\end{align}
We also find analytically the following normalization constants over the subsets $\mathcal{S}$ and $\mathcal{O}_S$:
\begin{align}
    \mu(\strat\in\mathcal{S})=\frac{30}{81},&& p(\outa\in\mathcal{O}_S)=\frac{30}{81}.
\end{align}
We find they are equal, as expected. From Eqn.~\eqref{eqn:q_lambda}, we find
\begin{align}
    q(\strat)=\frac{\mu(\strat)}{\mu(\strat\in\mathcal{S})}=\frac{1}{30},&& \forall \strat\in\mathcal{S}.
\end{align}

\subsection*{Step 3b: Constraint Class 4}
Since $q(\strat)$ is uniformly distributed, the pre-factors $1/\mu(\strat)$ and $1/\mu(\strat')$ in Eqn.~\eqref{eq:themarg} cancel. Each side of the equivalence constraint corresponds to 8 decision variables (i.e. the fine-graining of 3 single-click outcomes with $a_n=\xx{X}$ fixed as $\xx{L}$ or $\xx{R}$).

\subsection*{Step 3c: Constraint Class 5}
We compute the domain asymmetries analytically using Eqn.~\eqref{eq:58}. For all $n$ and $t$, we find
\begin{align}
    \Delta_{\mathcal{O}_{a_n=\xx{L}}}&=(t-1/2)/15,\\
    \Delta_{\mathcal{O}_{a_n=\xx{R}}}&=(1/2-t)/15.
\end{align}
We then relate them to $q(\outa,\strat)$ via Eqn.~\eqref{eqn:ring_network_domain_asymmetry_constraints}.

\subsection*{Workings for $\mathcal{O}_{a_1=\xx{L}}$}
Consider the example $a_1=\xx{L}$. We find the outcome pattern sets corresponding to $\outa^{(1,1)}$ and $\outa^{(4,1)}$ are:
\begin{align}
    \mathcal{O}_p^{(1)}=\{\xx{L00XXX}\}, && \mathcal{O}_p^{(2)}=\{\xx{LXXX00}\},\\
    \mathcal{S}_p^{(1)}=\{\strat_{10},\strat_{11}\}, && \mathcal{S}_p^{(2)}=\{\strat_{28},\strat_{29}\}.
\end{align}
Thus we compute the domain asymmetry as,
\begin{align}\label{eqn:6p4s_domain_asymmetry}
    \Delta_{\mathcal{O}_{a_1=\xx{L}}}&=\frac{p(\xx{L00XXX})-p(\xx{LXXX00})}{p(\outa\in\mathcal{O}_S)},
\end{align}
where we use $\xx{X}$ as a shorthand to mean the marginal over single-click outcomes. The domain asymmetry constraint is
\eqalign{
    \Delta_{\mathcal{O}_{a_1=\xx{L}}}=q(\outa^{(1,1)},\strat_{10})+q(\outa^{(1,1)},\strat_{11})\hspace{4em}\\
    -[q(\outa^{(4,1)},\strat_{28})+q(\outa^{(4,1)},\strat_{29})],
}
where each $q(\outa^{(m,1)},\strat_j)$ represents 8 decision variables. The source independence constraint for one strategy is
\begin{align}
    q(\strat_{10})=\frac{1}{30}=q(\xx{X00XXX},\strat_{10}),
\end{align}
which corresponds to the sum over 16 decision variables. We find the strategies $\strat_{10}$, $\strat_{16}$, $\strat_{18}$, $\strat_{22}$, and $\strat_{26}$ also produce the same conditional probability on $a_n=\xx{X}$, allowing us to write constraints of the form in Eqn.~\eqref{eq:themarg}. The above steps can be repeated for every party $A_n$ and for all strategies $\strat_j$. Constraint classes 1 and 2 given in Eqns.~\eqref{eqn:constraints1} and \eqref{eqn:constraints2} are straightforward to compute. Altogether, this gives us the LP for $q(\outa,\strat)$.

\subsection*{Results}

\begin{figure}
    \centering
    \includegraphics[width=\linewidth]{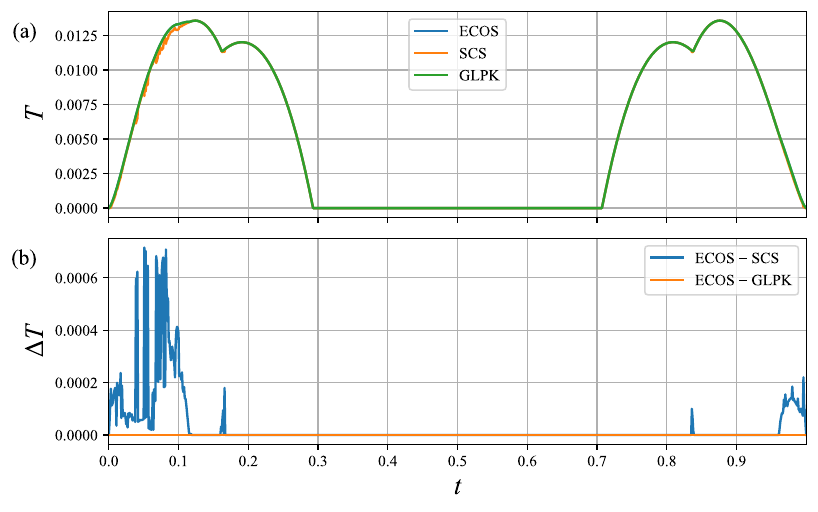}
    \caption{Numerical witness of network nonlocality. $T> 0$ corresponds to violation of network-locality conditions. (a) Minimized sum of tolerances against transmissivity for step size $\delta t=0.001$, for the 6-party, 4-source ring network in Fig.~\ref{fig:6p4s_ring_network}. (b) Difference in sum of tolerances between the ECOS solver, and SCS and GLPK solvers.}
    \label{fig:6p4s_LP_plot}
\end{figure}

As this is a feasibility witness, infeasibility of the LP is sufficient to certify network nonlocality. For numerical insight, we introduce non-negative tolerances on each constraint and set the linear program objective to minimize the sum of these tolerances, $T\equiv \sum \text{tols}$. Then $T$ can be thought of as an estimate of how much the constraints must be relaxed to match the observed statistics. For any value of $T>0$, the program is infeasible. We compare the results using ECOS, SCS, and GLPK solvers. For ECOS, we used an absolute tolerance, relative tolerance, and feasibility tolerance of $10^{-8}$, with a maximum of 500 iterations. For SCS, we used an overall convergence accuracy of $10^{-8}$, with a maximum of 20000 iterations. For GLPK, we used default settings. We find infeasibility and thus network nonlocality for approximately $t\in(0,0.292)\cup(0.708,1)$, for a step size of $\delta t=0.001$, as shown in Fig.~\ref{fig:6p4s_LP_plot}(a). Note the specific shape of the curve has no direct physical interpretation. To compare the solvers, we compare the difference in the sum of their tolerances using the ECOS solver as a reference in Fig.~\ref{fig:6p4s_LP_plot}(b). The GLPK and ECOS solvers differ in magnitude by $10^{-10}$ while SCS and ECOS differ in magnitude by $10^{-4}$. This is expected as the SCS is a first-order solver and ECOS is a second-order solver \cite{Domahidi2013ECOS, ODonoghue2016SCS}.

Since this witness can only prove the existence of network nonlocality for a given $t$, we cannot conclude anything about the intervals of $t$ for which the constraints are satisfied. That being said, the program is feasible for $t=0$ and $t=1$, when the beam splitter does not act as an entangling measurement, which is expected.

\subsection*{Numerical Analysis}
Several numerical methods exist for detecting network nonlocality, including variable-elimination, classical inflation, and rigidity-based arguments \cite{tavakoli2022bell}. While none of these methods work in full generality, each has their own strengths and weaknesses.

Variable-elimination alone encompasses a range of methods. One common thread is the use of conditional independence constraints, which occur whenever at least two parties do not receive the same LVs in the network. This has been used to derive nonlinear inequalities, such as the bilocal inequality, which leverages the conditional independence between $A$ and $C$, as shown in Fig.~\ref{fig:bilocal_network}(b) \cite{branciard2010characterizing}.

Classical inflation offers convex outer approximations of the set of network-local behavior by leveraging the no-cloning of quantum information and cloning of classical local variables. At inflation order $l$, a larger network is constructed with $l$ copies of each LV, and each party is replaced by clones corresponding to all combinations of LVs it can receive. If party $A_i$ depends on $k_i$ LVs and has $d$ possible outputs (assuming all parties have the same dimension of outputs), then in an order-$l$ inflation it becomes $l^{k_i}$ nodes. The inflated joint distribution then has
\begin{align}\label{eqn:inflation_scaling}
    d^{\sum_{i=1}^N l^{k_i}}
\end{align}
decision variables, which must satisfy various constraints including permutation symmetries. The number of permutation symmetry constraints grows at worst as $l!$. Thus inflation methods become computationally intractable beyond a few levels of the hierarchy \cite{Navascus2020inflation}.

Lastly, rigidity-based arguments are usually tailored for a specific network. If a network correlation is only compatible with a unique classical strategy (up to relabeling), it is called rigid. This rigidity can be used to derive constraints which can be proved to be incompatible with the observed distribution. It can be very powerful, as in the triangle network scenario, but can only be applied if rigidity is proven, which is not true for general networks \cite{Renou2022rigidity}.

Our method would fit within variable-elimination methods, as we explicitly enumerate an LV distribution $\mu(\strat)$ and test its compatibility with the observed statistics. We can use conditional independence and network properties to derive constraints on a restricted distribution $q(\strat)$. Specifically, we derive strategy distribution, conditional independence, and domain asymmetry constraints. The number of decision variables in our linear program will depend on the number of observed outcomes in the restricted set, $|\mathcal{O}_S|$, as well as the number of local strategies in its pre-image, $|\mathcal{S}|$. The size of the restricted output set is upper bounded $|\mathcal{O}_S|< d^N$, as $d^N$ is the maximum size of $\mathcal{O}$ and does not account for unphysical outcomes (e.g. which don't satisfy photon-number-conservation). For a network with $M$ sources distributing to $P$ parties, $|\mathcal{S}|< P^M$, as $\mathcal{S}\subset \mathcal{D}$. Since $P$ is fixed (and likely small), this will usually scale better than Eqn.~\eqref{eqn:inflation_scaling}.

Note that this scaling is also better than the usual enumeration of deterministic strategies discussed in Sec.~\ref{sec:background}. For a network with $d_{x_i}$-dimensional sets of input questions and $d_{a_i}$-dimensional sets of output answers, the number of strategies scales as
\begin{align}
    \prod_{i=1}^N d_{a_i}^{d_{x_i}}.
\end{align}
Additionally, our method also does not rely on finding unique classical strategies as in the rigidity arguments, making it more versatile.

\section{Conclusion}\label{sec:conclusion}
We now remark on our procedure and situate our results within more specific definitions of network nonlocality.

Observe that our choice of source-to-party connections means each party receives a unique LV pair. This configuration results in exactly two strategies for an outcome with fixed $\xx{0}$ detections in the subset $\mathcal{O}_S$. However, this choice is not necessary to compute valid domain asymmetries. In fact, for a network configuration where multiple parties receive the same LV pair, using the symmetries of the network to group strategies will collapse the number of linearly independent constraints to the same as the scenario with unique LV pairs.

The standard definition of network-locality, as in Eqn.~\eqref{eqn:network_local_cond}, does not distinguish whether nonlocality arises at a single source or is distributed across a network. This has motivated definitions for network nonlocality which inherently require a network structure, called \textit{full network nonlocality} and \textit{genuine network nonlocality} \cite{pozas2022full, vsupic2022genuine}.

From an intuitive perspective, full network nonlocality can be thought of as requiring every source in a network to generate nonlocal correlations (e.g. emit entangled states), while genuine network nonlocality can be thought of as requiring at least one measurement be entangling (e.g. a Bell-state measurement). For our system, sources distribute tripartite W states which provide entangled resources. Furthermore, when $t\in(0,1)$, the beamsplitter acts as an entangling measurement on the two optical modes. Both these suggest our system is consistent with the definitions of full network nonlocality and genuine network nonlocality, although a proper certification would require ruling out models with a mix of local and nonlocal network elements.

In conclusion, we have presented five sufficient classes of linear constraints on an auxiliary distribution related to the full statistics of a system which form a witness for network nonlocality for arbitrary networks. We outlined the general procedure for deriving the constraint classes, including the interpretation of the domain constraints. Using our methodology, we certified network nonlocality for a 6-party ring network with 4 sources. The fact that there are fewer photons than parties in the system allowed us to infer the behavior of each LV over the restricted subset $\mathcal{O}_S$. Our approach provides a framework to search for LP witnesses of network nonlocality in other networks.

\section*{Acknowledgments}
This work was supported by the India-Japan Cooperative Science Programme under Grant 120257718. This work is partially carried out under IRA Programme, project no.~FENG.02.01-IP.05-0006/23, financed by the FENG program 2021-2027, Priority FENG.02, Measure FENG.02.01., with the support of the FNP. This research was partially conducted while visiting the Okinawa Institute of Science and Technology (OIST) through the Theoretical Sciences Visiting Program (TSVP).

\bibliographystyle{IEEEtran}
\bibliography{reference}


{\appendices
\begin{table}
    \centering
    \caption{Detection patterns and their corresponding strategies for the network in Fig.~\ref{fig:6p4s_ring_network}. $\lambda_m=A_n$ means LV $\lambda_m$ sends a photon to party $A_n$.}
    \begin{tabular}{ccccc c}
        \toprule
        $\lambda_1$ & $\lambda_2$ & $\lambda_3$ & $\lambda_4$ & \text{Strategy Label} & $F(\strat)\mapsto \mathcal{O}_S$ \\
        \midrule
        $A_3$ & $A_4$ & $A_6$ & $A_5$ & $\strat_{0}$  & \texttt{00XXXX} \\
        $A_3$ & $A_5$ & $A_4$ & $A_6$ & $\strat_{1}$  & \texttt{00XXXX} \\ \midrule

        $A_2$ & $A_4$ & $A_6$ & $A_5$ & $\strat_{2}$  & \texttt{0X0XXX} \\
        $A_2$ & $A_5$ & $A_4$ & $A_6$ & $\strat_{3}$  & \texttt{0X0XXX} \\ \midrule

        $A_2$ & $A_5$ & $A_3$ & $A_6$ & $\strat_{4}$  & \texttt{0XX0XX} \\
        $A_3$ & $A_2$ & $A_6$ & $A_5$ & $\strat_{5}$  & \texttt{0XX0XX} \\ \midrule

        $A_2$ & $A_4$ & $A_3$ & $A_6$ & $\strat_{6}$  & \texttt{0XXX0X} \\
        $A_3$ & $A_2$ & $A_4$ & $A_6$ & $\strat_{7}$  & \texttt{0XXX0X} \\ \midrule

        $A_2$ & $A_4$ & $A_3$ & $A_5$ & $\strat_{8}$  & \texttt{0XXXX0} \\
        $A_3$ & $A_2$ & $A_4$ & $A_5$ & $\strat_{9}$  & \texttt{0XXXX0} \\ \midrule

        $A_1$ & $A_4$ & $A_6$ & $A_5$ & $\strat_{10}$ & \texttt{X00XXX} \\
        $A_1$ & $A_5$ & $A_4$ & $A_6$ & $\strat_{11}$ & \texttt{X00XXX} \\ \midrule

        $A_1$ & $A_5$ & $A_3$ & $A_6$ & $\strat_{12}$ & \texttt{X0X0XX} \\
        $A_3$ & $A_5$ & $A_6$ & $A_1$ & $\strat_{13}$ & \texttt{X0X0XX} \\ \midrule

        $A_1$ & $A_4$ & $A_3$ & $A_6$ & $\strat_{14}$ & \texttt{X0XX0X} \\
        $A_3$ & $A_4$ & $A_6$ & $A_1$ & $\strat_{15}$ & \texttt{X0XX0X} \\ \midrule

        $A_1$ & $A_4$ & $A_3$ & $A_5$ & $\strat_{16}$ & \texttt{X0XXX0} \\
        $A_3$ & $A_5$ & $A_4$ & $A_1$ & $\strat_{17}$ & \texttt{X0XXX0} \\ \midrule

        $A_1$ & $A_2$ & $A_6$ & $A_5$ & $\strat_{18}$ & \texttt{XX00XX} \\
        $A_2$ & $A_5$ & $A_6$ & $A_1$ & $\strat_{19}$ & \texttt{XX00XX} \\ \midrule

        $A_1$ & $A_2$ & $A_4$ & $A_6$ & $\strat_{20}$ & \texttt{XX0X0X} \\
        $A_2$ & $A_4$ & $A_6$ & $A_1$ & $\strat_{21}$ & \texttt{XX0X0X} \\ \midrule

        $A_1$ & $A_2$ & $A_4$ & $A_5$ & $\strat_{22}$ & \texttt{XX0XX0} \\
        $A_2$ & $A_5$ & $A_4$ & $A_1$ & $\strat_{23}$ & \texttt{XX0XX0} \\ \midrule

        $A_1$ & $A_2$ & $A_3$ & $A_6$ & $\strat_{24}$ & \texttt{XXX00X} \\
        $A_3$ & $A_2$ & $A_6$ & $A_1$ & $\strat_{25}$ & \texttt{XXX00X} \\ \midrule

        $A_1$ & $A_2$ & $A_3$ & $A_5$ & $\strat_{26}$ & \texttt{XXX0X0} \\
        $A_2$ & $A_5$ & $A_3$ & $A_1$ & $\strat_{27}$ & \texttt{XXX0X0} \\ \midrule

        $A_2$ & $A_4$ & $A_3$ & $A_1$ & $\strat_{28}$ & \texttt{XXXX00} \\
        $A_3$ & $A_2$ & $A_4$ & $A_1$ & $\strat_{29}$ & \texttt{XXXX00} \\
        \bottomrule
    \end{tabular}
    \label{table:6p4s_ring_lv_strats}
\end{table}

\section{Proof of Theorem~\ref{thm:lv_value}}\label{appendix:proof_lv_value_thm}
We first claim the set of outcomes in $\mathcal{O}_S$ compatible with $\outa^{(m,n)}$ are different from the set of outcomes in $\mathcal{O}_S$ compatible with $\outa^{(m',n)}$. Notice in $\mathcal{O}_S$ the local outcome $a_n=\xx{X}$ can occur if either ``$A_n$ receives a photon from $\lambda_m$" or ``$A_n$ receives a photon from $\lambda_{m'}$". Due to Prop.~\ref{prop:S_rules}, it cannot happen that both $\lambda_m$ and $\lambda_{m'}$ send their photon to $A_n$. Therefore, due to photon number conservation and Def.~\ref{def:strategy}, all parties in the joint target list minus $A_n$ cannot simultaneously be $\xx{0}$.

Now $\outa^{(m,n)}$ is compatible with all outcomes that have $(a_u,a_v)=\xx{0}$ and hence at least one of the remaining parties in $(a_{u'},a_{v'})\neq \xx{0}$. Therefore, these outcomes are not compatible with the substring $\outa^{(m',n)}$. Similarly, $\outa^{(m',n)}$ is compatible with all outcomes that have at least one of the remaining parties $(a_u,a_v)\neq \xx{0}$, and hence are not compatible with the substring $\outa^{(m,n)}$. This concludes our claim. Note this holds for $P$-partite single-photon sources where $\lambda_m$ and $\lambda_{m'}$ can share as few as one target party $A_n$, or as many as $P-1$ target parties.

With the fact that $\outa^{(m,n)}$ and $\outa^{(m',n)}$ correspond to mutually exclusive events in $\mathcal{O}_S$ established, we prove Eqn.~\eqref{eqn:thm1_claim1}. The proof for Eqn.~\eqref{eqn:thm1_claim2} is analogous.

Consider an outcome $\outa\in\mathcal{O}_S$ compatible with the substring $\outa^{(m,n)}$. By Prop.~\ref{prop:S_rules}, it cannot receive a photon from both $\lambda_m$ and $\lambda_{m'}$. Suppose $\lambda_m\neq A_n$. Due to photon number conservation and Def.~\ref{def:strategy}, $\lambda_m$ must send exactly one photon to one of the remaining parties in its target list $\{A_u,A_v\}$, meaning at least one of them must output $\xx{X}$. However, by Def.~\ref{def:outcome_substring}, the remaining parties $(a_u,a_v)=\xx{0}$. We have a contradiction. Thus $\lambda_m=A_n$. This concludes the proof of Eqn.~\eqref{eqn:thm1_claim1}.

\section{Proof of Theorem~\ref{thm:prob_lv_subset}}\label{appendix:proof_thm2}
To verify Eqns.~\eqref{eqn:sp1} and \eqref{eqn:sp2} are correct, recall from Thm.~\ref{thm:lv_value}, if an outcome $\outa\in\mathcal{O}_S$ is compatible with the substring $\outa^{(m,n)}$, we know the value of one of the LVs is $\lambda_m=A_n$. All strategies in the support of $\outa$ must be in $\mathcal{S}$ and from Prop.~\ref{prop:S_rules}, strategies in $\mathcal{S}$ cannot have $\lambda_m=\lambda_{m'}=A_n$. Therefore, $\lambda_{m'}\neq A_n$. Furthermore, since $(a_u,a_v)=\xx{0}$, we know the other LVs cannot send a photon to any party in the target list of $\lambda_m$, meaning $\stratrest{m}\neq\{A_u,A_v,A_w\}$. Hence, the strategies in the support of $\mathcal{O}_p^{(1)}$ must satisfy
\eqalign{\label{eqn:sp1_proof}
    \mathcal{S}_p^{(1)}=\{\strat&\in\mathcal{S}:\\
    &\lambda_m=A_n,\stratrest{m}\neq \{A_u,A_v,A_w\}\},
}
A similar argument can be made for $\mathcal{S}_p^{(2)}$.

To verify Eqn.~\ref{eqn:weight_sp1} is correct, consider $\mathcal{O}_p^{(1)}$, the set of outcome patterns compatible with the substring $\outa^{(m,n)}$, as defined in Eqn.~\ref{eqn:op1}. We have the following
\begin{align}
    \mu\left(\outa^{(m,n)}\right)&=\mu\left(\outa\in\mathcal{O}_p^{(1)}\right)\\
    &=\sum_{\bar{\outa}\in\mathcal{O}_p^{(1)}}\mu(\bar{\outa})\\
    &=\sum_{\strat\in\mathcal{S}_p^{(1)}} \sum_{\bar{\outa}\in\mathcal{O}_p^{(1)}} \mu(\bar{\outa},\strat) \label{eqn:thm2_proof_line2}\\
    &=\sum_{\strat\in\mathcal{S}_p^{(1)}} \mu(\strat) \label{eqn:thm2_proof_line3}\\
    &=\mu\left(\strat\in\mathcal{S}_p^{(1)}\right).
\end{align}
In Eqn.~\eqref{eqn:thm2_proof_line2}, we use Prop.~\ref{prop:f_sp_op}. This allows us to quantify the total weight of the LV subset $\mathcal{S}_p^{(1)}$. A similar argument holds for $\mu(\outa^{(m',n)})$.

To verify Eqn.~\ref{eqn:weight_sp1_q} is correct, observe we get the following by dividing across by the normalizations over $\mathcal{S}$ and $\mathcal{O}_S$ in Eqn.~\ref{eqn:weight_sp1},
\begin{align}
    \frac{\mu(\strat\in\mathcal{S}_p^{(1)})}{\mu(\strat\in\mathcal{S})\mu(\outa\in\mathcal{O}_S)}&=\frac{\mu(\outa\in\mathcal{O}_p^{(1)})}{\mu(\strat\in\mathcal{S})\mu(\outa\in\mathcal{O}_S)},\\
    q(\strat\in\mathcal{S}_p^{(1)})&=\frac{\mu(\outa\in\mathcal{O}_S)}{\mu(\strat\in\mathcal{S})}q(\outa\in\mathcal{O}_p^{(1)}).\label{eqn:qS_qO_ratio}
\end{align}
We now claim $\mu(\outa\in\mathcal{O}_S)= \mu(\strat\in\mathcal{S})$. To show this, first observe
\begin{align}
    q\left(\strat\in\mathcal{S}_p^{(1)}\right) =q\left(\outa\in\mathcal{O}_p^{(1)},\strat\in\mathcal{S}_p^{(1)}\right),\label{eqn:qS_qjoint}
\end{align}
since $F(\mathcal{S}_p)=\mathcal{O}_p$ by Prop.~\ref{prop:f_sp_op}. Furthermore, observe
\begin{align}
    q\left(\outa\in\mathcal{O}_p^{(1)}\right)& =\sum_{\outa\in\mathcal{O}_p^{(1)}}\sum_{\strat\in\mathcal{S}_p^{(1)}}q(\outa,\strat),\\
    &=q\left(\outa\in\mathcal{O}_p^{(1)},\strat\in\mathcal{S}_p^{(1)}\right).\label{eqn:qO_qjoint}
\end{align}
by the definition in Eqn.~\eqref{eqn:qa_marginal} and $F^{-1}(\mathcal{O}_p)=\mathcal{S}_p$. Combining Eqns.~\eqref{eqn:qS_qjoint} and \eqref{eqn:qO_qjoint}, we have
\begin{align}
    q\left(\strat\in\mathcal{S}_p^{(1)}\right)&= q\left(\outa\in\mathcal{O}_p^{(1)}\right)\\
    &=\frac{\mu(\outa\in\mathcal{O}_S)}{\mu(\strat\in\mathcal{S})}q(\outa\in\mathcal{O}_p^{(1)}),
\end{align}
where we have used Eqn.~\eqref{eqn:qS_qO_ratio} in the second line. When $q(\strat\in\mathcal{S}_p^{(1)}),\: q(\outa\in\mathcal{O}_p^{(1)})>0$, then $\mu(\outa\in\mathcal{O}_S)=\mu(\strat\in\mathcal{S})$. This concludes our claim.
}

\end{document}